\documentstyle[12pt]{article}

\catcode`\@=11

\@addtoreset{equation}{section}
\def\theequation{\arabic{section}.\arabic{equation}}

\def\@normalsize{\@setsize\normalsize{15pt}\xiipt\@xiipt
\abovedisplayskip 14pt plus3pt minus3pt%
\belowdisplayskip \abovedisplayskip
\abovedisplayshortskip  \z@ plus3pt%
\belowdisplayshortskip  7pt plus3.5pt minus0pt}

\def\small{\@setsize\small{13.6pt}\xipt\@xipt
\abovedisplayskip 13pt plus3pt minus3pt%
\belowdisplayskip \abovedisplayskip
\abovedisplayshortskip  \z@ plus3pt%
\belowdisplayshortskip  7pt plus3.5pt minus0pt
\def\@listi{\parsep 4.5pt plus 2pt minus 1pt
            \itemsep \parsep
            \topsep 9pt plus 3pt minus 3pt}}

\def\underline#1{\relax\ifmmode\@@underline#1\else
        $\@@underline{\hbox{#1}}$\relax\fi}
\@twosidetrue

\relax

\catcode`@=12

\catcode`\@=11

\def\section{\@startsection{section}{1}{\z@}{3.5ex plus 1ex minus
   .2ex}{2.3ex plus .2ex}{\large\bf}}

\def\thesection{\Roman{section}.}

\def\appendix{\setcounter{section}{0}
        \def\thesection{Appendix }
        \def\theequation{\Alph{section}.\arabic{equation}}}

\def\ps@headings{\def\@oddfoot{}\def\@evenfoot{}
\def\@oddhead{\hbox{}\hfill
        \makebox[.5\textwidth]{\raggedright\ignorespaces --\thepage{}--
        \hfill {}}}
\def\@oddhead{\hbox{}\hfill --\thepage{}-- \hfill
        {}}
\def\@evenhead{\@oddhead}
\def\subsectionmark##1{\markboth{##1}{}}
}

\ps@headings

\catcode`\@=12

\relax

\def\figcap{\section*{Figure Captions\markboth
        {FIGURECAPTIONS}{FIGURECAPTIONS}}\list
        {Fig. \arabic{enumi}:\hfill}{\settowidth\labelwidth{Fig. 999:}
        \leftmargin\labelwidth
        \advance\leftmargin\labelsep\usecounter{enumi}}}
 \relax
\def\tablecap{\section*{Table Captions\markboth
        {TABLECAPTIONS}{TABLECAPTIONS}}\list
        {Table \arabic{enumi}:\hfill}{\settowidth\labelwidth{Table 999:}
        \leftmargin\labelwidth
        \advance\leftmargin\labelsep\usecounter{enumi}}}
 \relax
\def\reflist{\section*{References\markboth
        {REFLIST}{REFLIST}}\list
        {[\arabic{enumi}]\hfill}{\settowidth\labelwidth{[999]}
        \leftmargin\labelwidth
        \advance\leftmargin\labelsep\usecounter{enumi}}}
 \relax

\catcode`\@=11

\def\ps@headings{\def\@oddfoot{}\def\@evenfoot{}
\def\@oddhead{\hbox{}\hfill
        \makebox[.5\textwidth]{\raggedright\ignorespaces --\thepage{}--
        \hfill {}}}
\def\@evenhead{\@oddhead}
\def\subsectionmark##1{\markboth{##1}{}}
}

\ps@headings

\relax

\catcode`\@=11

\@addtoreset{equation}{section}
\def\theequation{\arabic{section}.\arabic{equation}}

\def\@normalsize{\@setsize\normalsize{15pt}\xiipt\@xiipt
\abovedisplayskip 14pt plus3pt minus3pt%
\belowdisplayskip \abovedisplayskip
\abovedisplayshortskip  \z@ plus3pt%
\belowdisplayshortskip  7pt plus3.5pt minus0pt}

\def\small{\@setsize\small{13.6pt}\xipt\@xipt
\abovedisplayskip 13pt plus3pt minus3pt%
\belowdisplayskip \abovedisplayskip
\abovedisplayshortskip  \z@ plus3pt%
\belowdisplayshortskip  7pt plus3.5pt minus0pt
\def\@listi{\parsep 4.5pt plus 2pt minus 1pt
            \itemsep \parsep
            \topsep 9pt plus 3pt minus 3pt}}

\def\underline#1{\relax\ifmmode\@@underline#1\else
        $\@@underline{\hbox{#1}}$\relax\fi}
\@twosidetrue

\relax

\catcode`@=12

\catcode`\@=11

\def\section{\@startsection{section}{1}{\z@}{3.5ex plus 1ex minus
   .2ex}{2.3ex plus .2ex}{\large\bf}}

\def\thesection{\Roman{section}.}

\def\appendix{\setcounter{section}{0}
        \def\thesection{Appendix }
        \def\theequation{\Alph{section}.\arabic{equation}}}

\def\ps@headings{\def\@oddfoot{}\def\@evenfoot{}
\def\@oddhead{\hbox{}\hfill
        \makebox[.5\textwidth]{\raggedright\ignorespaces --\thepage{}--
        \hfill {}}}
\def\@oddhead{\hbox{}\hfill --\thepage{}-- \hfill
        {}}
\def\@evenhead{\@oddhead}
\def\subsectionmark##1{\markboth{##1}{}}
}

\ps@headings

\catcode`\@=12

\relax

\def\figcap{\section*{Figure Captions\markboth
        {FIGURECAPTIONS}{FIGURECAPTIONS}}\list
        {Fig. \arabic{enumi}:\hfill}{\settowidth\labelwidth{Fig. 999:}
        \leftmargin\labelwidth
        \advance\leftmargin\labelsep\usecounter{enumi}}}
 \relax
\def\tablecap{\section*{Table Captions\markboth
        {TABLECAPTIONS}{TABLECAPTIONS}}\list
        {Table \arabic{enumi}:\hfill}{\settowidth\labelwidth{Table 999:}
        \leftmargin\labelwidth
        \advance\leftmargin\labelsep\usecounter{enumi}}}
 \relax
\def\reflist{\section*{References\markboth
        {REFLIST}{REFLIST}}\list
        {[\arabic{enumi}]\hfill}{\settowidth\labelwidth{[999]}
        \leftmargin\labelwidth
        \advance\leftmargin\labelsep\usecounter{enumi}}}
 \relax

\catcode`\@=11

\def\ps@headings{\def\@oddfoot{}\def\@evenfoot{}
\def\@oddhead{\hbox{}\hfill
        \makebox[.5\textwidth]{\raggedright\ignorespaces --\thepage{}--
        \hfill {}}}
\def\@evenhead{\@oddhead}
\def\subsectionmark##1{\markboth{##1}{}}
}

\ps@headings

\relax

\newskip\humongous \humongous=0pt plus 1000pt minus 1000pt

\newif\ifdtup

\def\beq{\begin{equation}}
\def\eeq{\end{equation}}

\def\beqn{\begin{eqnarray}}
\def\eeqn{\end{eqnarray}}
\relax

\def\G2{{\; \rm GeV/}c^2}
\def\G{\; \rm GeV}

\def\dotx{\dotx{\dot\overline{x}}}

\relax

\input {epsf}

\newbox \JCCHoldBox
\newdimen \JCCLower
\newcommand \epscenterbox [2]%
{    
     \epsfxsize=#1%
     \setbox \JCCHoldBox = \hbox{\epsfbox{#2}}%
     \JCCLower = 0.5\ht\JCCHoldBox
     \advance \JCCLower by -0.3ex%
     \lower \JCCLower \box\JCCHoldBox
}
\newcommand \epscenterboxy [2]%
{    
     \epsfysize=#1%
     \setbox \JCCHoldBox = \hbox{\epsfbox{#2}}%
     \JCCLower = 0.5\ht\JCCHoldBox
     \advance \JCCLower by -0.3ex%
     \lower \JCCLower \box\JCCHoldBox
}

\def\p{\partial}

\def\Lm{\Lambda}

\begin{document}
\begin{titlepage}
\begin{flushright}
       {\normalsize  hep-th 9511220 \\
                   OU-HET 222  \\
                November,~1995  \\}
\end{flushright}
\vfill
\begin{center}
  {\large \bf
 Macroscopic $n$-Loop Amplitude  for  Minimal Models Coupled
  to Two-Dimensional Gravity  \\
   ---  Fusion Rules and  Interactions --- }
\footnote{This work is supported in part by
  Grant-in-Aid for  Scientific Research
(07640403)
and by the Grand-in-Aid for Scientific Research Fund
(2690)
from
 the Ministry of Education, Science and Culture, Japan.}
\vfill
           {\bf M. Anazawa}$~^{\dag}$ \footnote{JSPS fellow}
 \\
                          and \\
         {\bf H.~Itoyama}$~^{\dag}$  \\
\vfill
       $~^{\dag}$ Department of Physics,\\
        Graduate School of Science, Osaka University,\\
        Toyonaka, Osaka, 560 Japan\\

\end{center}
\vfill
\begin{abstract}
 We investigate  the structure of  the macroscopic $n$-loop amplitude
 obtained from the two-matrix model at the unitary minimal critical
  point $(m+1,m)$.
 We  derive a general formula for  the $n$-resolvent  correlator at
 the continuum planar limit whose inverse Laplace transform provides
 the amplitude in terms of  the boundary lengths $\ell_{i}$
  and the renormalized cosmological constant $t$.
  The amplitude  is found to  contain a term  consisting of
 $\left( \frac{\partial}
  {\partial t} \right)^{n-3}$ multiplied by the product of modified
 Bessel functions summed over their degrees  which  conform to
  the fusion rules  and  the crossing symmetry.
   This is  found to be supplemented by  an increasing number of
 other terms with $n$
 which represent residual interactions  of loops.
  We reveal the nature of these
 interactions  by explicitly determining  them
 as the convolution of modified Bessel functions  and their
  derivatives  for the case $n=4$  and  the case $n=5$.
 We derive   a set of recursion relations
   which  relate     the terms in the $n$-resolvents  to those
  in   the $(n-1)$-resolvents.
\end{abstract}
\vfill
\end{titlepage}

\section{ Introduction and Conclusion}

  Matrix models provide an arena in which   the notion of
 integrability   is realized  as noncritical  string theory.
  At the same time, they produce efficient computation of
 some quantities  which would be very formidable in   the continuum
 framework.   Computation of macroscopic loop amplitudes\cite{MSS,MS,NB}
\cite{AIMZ}
  demonstrates   this fact most explicitly:  the boundary condition
  which is hard to solve in   the continuum framework \cite{NKYM}
  turns out to be related to the most natural quantity  in matrix models.
  Let us begin with recalling  this.

    A crude correspondence of  matrix models with   path integrals
  of noncritical strings    tells us that
   the connected part of  the correlator   given
  by  averaging over matrix  integrals of the product of singlet correlators
\beqn
  << tr \hat{M}^{q_{1}} tr \hat{M}^{q_{2}} \cdots
 tr \hat{M}^{q_{n}} >>_{N,conn}
\eeqn
   is an n-punctured surface swept by a noncritical string.
   To  turn  these punctures into holes  of a macroscpic size,
  one first introduces
  a  fixed loop length at the  i-th  boundary by  $\ell_{i}=  a q_{i}$.
  We are
   naturally led  to consider  the limiting procedure
\beqn
  {\cal A}_{n}\left( \ell_{1}, \ell_{2}, \cdots \ell_{n} \right)
  \equiv
  \lim_{q_{i} \rightarrow \infty, a \rightarrow 0, }
   \lim_{ N \rightarrow \infty}
  \frac{1}{\kappa^{n-2}}
  << tr \hat{M}^{q_{1}} tr \hat{M}^{q_{2}}
 \cdots tr \hat{M}^{q_{n}} >>_{N, conn}
\eeqn
   which defines   the macroscopic $n$-loop
 amplitude.\footnote{ See later sections for more of the definitions.}
   Here  $\kappa$
   is the renormalized string coupling  and $a$ is an auxiliary
 parameter  which plays the role of a cutoff.

  An equivalent  and more efficient  procedure is  to consider
     the correlator consisting of  the product of $n$-resolvents
  $<< {\displaystyle \prod_{i=1}^{n} }
 tr \frac{1}{p_{i}- \hat{M}}  >>_{N, conn}$,
 to pick its most  singular piece  and  finally to carry out  the inverse
  Laplace transforms  over $p_{i}$'s.  This in turn  means
\beqn
\label{eq:procedure}
{\cal A}_{n}\left( \ell_{1}, \ell_{2}, \cdots \ell_{n} \right)
  = (\prod_{j=1}^{n} {\cal L}_{j}^{-1})
  \lim_{a\rightarrow 0} \lim_{ N \rightarrow \infty}
  \frac{a^{n}}{\kappa^{n-2}}
 << \prod_{i=1}^{n} tr \frac{1}{p_{i}- \hat{M}}  >>_{N, conn} \;\;\;.
\eeqn
   Here, ${\cal L}_{j}^{-1}$  denotes the inverse Laplace  transform
  with respect to  $\zeta_{i}$ such that $a \zeta_{i} = p_{i}- p_{i}^{*}$
  and   $p_{i}^{*}$ denotes the critical value of $p_{i}$.
  In this paper, we  will carry out this procedure in depth
 at the  $(m+1, m)$  critical point  realized by the symmetric
   potential  of the two-matrix model. \footnote{  For some of the recent
  works on the two-matrix  models, see, for instance, \cite{twomatrix}.}
  Here  $\kappa \equiv \frac{1}{N a^{2+1/m}}$.

  In the next section, we  evaluate  the connected  part of  the correlator
  consisting of the product of $n$-resolvents  for large $N$  just mentioned
  above
  and derive a general formula for this object in the continuum planar
  limit.  We exploit  the planar solution  to the Heisenberg algebra
  and its parametrization provided by \cite{DKK}.
   Our formula contains   a term distinguishable  from others,
 namely  the one which  is expressible as   the total
 $\left( \frac{\partial} {\partial t} \right)^{n-3}$
  derivatives. Here $t$  denotes the renormalized cosmological constant.
  This  structure is familiar from the
  case of pure two-dimensional gravity.   This term is, however,
   found to be supplemented, for $n\geq 4$, by an increasing number
  of other terms with $n$.  This latter structure  testifies to the  existence
  of  interactions  which cannot be  captured by the  naive notion
  of operator product expansion   for microscopic loop operators:
  the macroscopic loop operator  will be expanded  by these.
   For that reason,  these interactions may be referred  to as contact
 interactions.

    In section III,  we   consider   the
 $\left( \frac{\partial} {\partial t} \right)^{n-3}$ $\cdots$ term.
    We are successful in representing this terms as
     the summations over $2n-3$ indices with
    its summand in a form of $n$ factorized products.
 These summations are  found to conform to
  the fusion rules  and  the crossing symmetry for the dressed primaries
  of the unitary minimal conformal field theory \cite{BPZ}.
 Using the formula for the inverse Laplace transform  found in \cite{AII},
\beqn
\label{eq:inverse}
{\cal L}^{-1}[\frac{\partial}{\partial \zeta}
         \frac{\sinh k \theta}{\sinh m \theta}]
= - \frac{M \ell}{\pi} \sin \frac{k \pi}{m}~  K_{k/m} (M \ell)
 \equiv  - \frac{M \ell}{\pi} \underline{K}_{k/m} (M \ell)\;\;.
\eeqn
 we determine the complete form for this part
  of the amplitude in terms of the boundary lengths
  $\ell_{i}$ $i=1, \cdots, n$. The answer reads as
\beqn
 &&  {\cal A}_{n}^{ fusion}( \ell_{1}, \ell_{2}, \cdots \ell_{n})
  \nonumber \\
 &&\qquad = - \frac{1}{m} \left( \frac{1}{m+1} \right)^{n-2}
  \left( \frac{\partial}{\partial t} \right)^{n-3}
 \left[ t^{-1 - \frac{(n-2)}{2m} } \sum_{ {\cal D}_{n}}  \prod_{j=1}^{n}
  \frac{M\ell_{j}}{\pi} \underline{K}_{1-k_j/m } \left( M \ell_{j} \right)
 \right] \;\;.
\eeqn
  The case $n=3$  has been briefly reported  in \cite{AII2}.
 In section IV,  we  consider the remaining pieces in the formula
  which represent  the residual interactions  of loops.
 For the case  $n=4$  and the case $n=5$, we have succeeded in expressing
  these in terms of the convolution of modified Bessel functions  and their
  derivatives.
  We,  therefore, obtain the complete answer  for
${\cal A}_{4}\left( \ell_{1},  \cdots, \ell_{4} \right)$ and the one for
${\cal A}_{5}\left( \ell_{1},  \cdots, \ell_{5} \right)$, which
  are eq.~(\ref{eq:n=4answer}) and eq.~(\ref{eq:n=5answer})  respectively.
  Although it is not unlikely that
   one can    determine   the full amplitude this way for
  arbitrary $n$,    the proof
  remains elusive.    We will finish with a few remarks  concerning
 with    the properties of these residual interactions.

 In  Appendix A, we derive  a set of recursion relations  which
   are used to evaluate the formula in section III.
 These recursion relations relate the expression
  of the terms appearing in the $n$-resolvent
   to    those in  the $(n-1)$-resolvent.
 These define, therefore,    the $n$-loop amplitude in terms
  of $(n-1)$-loop amplitude through  the inverse Laplace transforms
 albeit being implicit.

\section{The $n$-Resolvent Correlator in  Continuum Planar  Limit}

  Consider in the two-matrix model  the connected  part of  the correlator
  consisting of the product of  $n$-resolvents at finite $N$:
\beq
\label{eq:nresolv}
   << tr\frac{1}{p_{1}- \hat{M}}  tr\frac{1}{p_{2}- \hat{M}}
  \cdots  tr\frac{1}{p_{n}- \hat{M}} >>_{N, conn} \;\;\;.
\eeq
  Here,  $(\hat{M}, \hat{\tilde{M}})$ are the matrix variables
   and $p_{i}$'s are  eigenvalue coordinates
  which, in the continuum limit,  become Laplace-conjugate  to  loop lengths.
  We denote by $<< \cdots >>_{N,conn}$  the averaging with respect to the
 matrix integrations.
It should be noted that
  this expression is at most $\left( \frac{1}{N}\right)^{n-2}$
    due to the large $N$
  factorization  of the correlator  consisting of
 the product of singlet operators.
  In the second quantized notation
 \footnote{ See, for example,
 \cite{BDSS,IM} },  eq.~(\ref{eq:nresolv}) is expressible as
\beqn
  &&{}_{N}< 0 \mid \prod_{i=1}^{n} : \int d \lambda_{i} b^{\dagger}
(\lambda_{i})
 \frac{1}{p_{i} - \lambda_{i}} b(\lambda_{i}) : \mid 0 >_{N}
\;\;\; \nonumber  \\
 &&\qquad = {}_{N}< 0 \mid \prod_{i=1}^{n} : B^{\dagger}_{k_{i}}B_{j_{i}} :
 \mid 0 >_{N}
  \prod_{i=1}^{n} < k_{i} \mid \frac{1}{p_{i} - \hat{M}} \mid j_{i} > \;\;\;,
\eeqn
  where
\beqn
 b(\lambda) &=& \sum_{j=0}^{\infty} <\lambda\mid j> B_{j} \;\;, \;\;\
 b^{\dagger} (\lambda) = \sum_{j=0}^{\infty}<k\mid \lambda> B_{k}^{\dagger}
  \;\;\;   \nonumber \\
  B_{j} \mid \Omega > &=& 0 \;\;,\;\;\; j= 0,1,2,
 \cdots \;\;\; \nonumber      \\   ~~~
 {\rm and}~~~~ \mid 0 >_{N} &\equiv& \prod_{j=0}^{N-1}
 B_{j}^{\dagger} \mid \Omega>\;\;\;.
\eeqn
  The normal ordering  $: ...:$ is with respect to
 the filled sea $\mid 0 >_{N}$.
  We introduce a notation
\beqn
  \left[ \frac{1}{p-\hat{M}} \right]\left( z_{i}; \Lambda_{i} ,
 \Lambda , N\right)
  \equiv \sum_{\delta} z_{i}^{\delta} < j_i -\delta\mid \frac{1}{p-\hat{M}}
 \mid j_i >
  \;\;\\
  \Lambda_{i}=  j_i \Lambda /N = \Lambda + \Lambda \tilde{j}_i /N \;\;\;.
\eeqn
 The evaluation of
${}_{N}< 0 \mid {\displaystyle \prod_{i=1}^{n} }
 : B^{\dagger}_{k_{i}}B_{j_{i}} : \mid 0 >_{N}$
  by the Wick theorem provides $(n-1)!$ terms
   of the  following structure:
 each  term is given by the product
 of $n$-Kronecker delta's multiplied  both by a sign factor   and
  by  the product of
  $n$-step functions to ensure that the summations over the $n$-indices
  $ \tilde{j}_{1}, \tilde{j}_{2} \cdots $  and  $\tilde{j}_{n}$ are   bounded
 either from below $(\geq 0)$ or from above$(\leq -1)$.
  We denote this product by $ \Theta( \tilde{j}_{1},
  \tilde{j}_{2}, \cdots \tilde{j}_{n}; \sigma)$.
  These $(n-1)!$ terms are in one-to-one correspondence
 with  the circular permutations of $n$ integers   $1, \cdots, n$, which
 we denote by ${\cal S}_{n}$.   The $\sigma$ is an element of ${\cal S}_{n}$.
 For large $N$, we find
\beqn
\label{eq:step1}
 &~& \left(\frac{N}{\Lambda}\right)^{n-2} << \prod_{i=1}^{n} tr
 \frac{1}{p_{i} - \hat{M}} >>_{N,conn}   \;\;\; \\
 &=& \sum_{\tilde{j_{1}}, \tilde{j_{2}}, \cdots \tilde{j_{n}}}
  \sum_{\sigma \in {\cal S}_{n}}  \Theta( \tilde{j}_{1},
  \tilde{j}_{2}, \cdots \tilde{j}_{n}; \sigma)  sgn (\sigma)
  \left(  \prod_{j=1}^{n} \oint
 \frac{dz_{j}}{2\pi i} \right)
  \prod_{k=1}^{n} \frac{1}{z_{k}}
 \left( \frac{z_{\sigma(k)}}{z_{k}} \right)^{\tilde{j}_{k}}  \nonumber  \\
 &~& \times \frac{1}{(n-2)!} \left( \sum_{i=1}^{n} \tilde{j}_{i}
 \frac{\partial}{ \partial \Lambda_{i} } \right)^{n-2}
 \prod_{i^{\prime}=1}^{n}  \frac{1}{p_{i^{\prime}} -
 \left[\hat{M}\right]  (z_{i^{\prime}}; \Lambda_{i^{\prime}})}
  \mid_{\Lambda_{i^{\prime}}
 = \Lambda}   + {\cal{O}}\left(1/N\right) \;\;\;.
\eeqn
Note that in the large $N$ limit, we can use $\frac{1}{p_i -
 \left[\hat{M}\right](z_i; \Lambda_i) }$
 in  place of  $\left[ \frac{1}{p_i - \hat{M}  }\right](z_i;\Lambda_i,
\Lambda,N)$
according to the same reason as stated in \cite{ AII2}.
Here the `classical' function is defined by
\beq
 \left[ \hat{M} \right] (z_i ; \Lambda_i)
 \equiv \lim_{N\to\infty} \sum_{\delta} z_{i}^{\delta} \left< j_i -\delta
 \right|\hat{M}\left|  j_i \right>
 \qquad , \;\;
 \Lambda_i = j_i \Lambda /N
 \quad .
\eeq
  The  $sgn(\sigma)$ denotes the signature associated with  the
 permutation $\sigma$.

   Let us define
\beq
  m! D_{m}(z, z^{\prime}) \equiv
 \frac{1}{z} \sum_{\tilde{j} \geq 0 } \tilde{j}^{m}
 \left(z^{\prime}/z\right)^{\tilde{j}}
  =  - \frac{1}{z} \sum_{\tilde{j} \leq -1 } \tilde{j}^{m}
 \left(z^{\prime}/z\right)^{\tilde{j}}  \;\;,  \;\;\; m= 0, \cdots \;\;\;.
\eeq
  In the continuum limit  we will be focusing  from now on,
 it is sufficient to use
\beq
  D_{m}(z, z^{\prime}) \approx \frac{1}{ (z- z^{\prime})^{m+1}}
  \equiv D_m(z-z') \quad .
\eeq
Let $sgn_{i}(\sigma)$ be $+1$ or $-1$, depending upon whether
  the restriction on the summation over $\tilde{j}_{i}$ is
  bounded from below or from above respectively.
  It is not difficult  to show
\beq
 sgn(\sigma) \prod_{i=1}^{n}sgn_{i}(\sigma) = -1 \;\;\;,
\eeq
  for any $\sigma$ and $n$.
 The summations over
 $ \tilde{j}_{1}, \tilde{j}_{2} \cdots $  and  $\tilde{j}_{n}$
  can then be  performed  for all $\sigma$  at once,
  leaving  with this minus sign.

  Now we turn to the integrations over $z_{i}$ $(i= 1 \sim n)$.
   The convergence on  the geometric series
 leads to  the  successively  ordered integrations of $z_{i}'s$
  for each $\sigma$.
 By  simply picking up a pole  of $z_{i}$ at
  $\frac{1}{p_{i} - M (z_{i} ; \Lambda_{i}) }$  for $i= 1 \sim n$  and
  using
\beqn
 &~&   \oint \frac{dz_{i}}{2\pi i} f\left( \cdots z_{i}, \cdots \right)
 \left( \frac{\partial}{\partial \Lambda_{i}} \right)^{\ell}
\left( \frac{1}{p_{i} -
  \left[ \hat{M} \right] (z_{i}; \Lambda_{i}) }\right)  \nonumber \\
 &=&  -  \frac{\partial}{\partial(a \zeta_{i})}
 \left( \frac{\partial}{\partial \Lambda_{i}} \right)^{\ell}
  \int^{z_{i}^{*}} dz_{i}  f\left( \cdots z_{i}, \cdots \right) \;\;\;,
 \;\; \ell =  0,1, \cdots \;\;\;,
\eeqn
  we find that eq.~(\ref{eq:step1}) is written as
\beqn
\label{eq:step2-1}
 \left(\frac{N}{\Lambda}\right)^{n-2} << \prod_{i=1}^{n} tr
 \frac{1}{p_{i} - \hat{M}} >>_{N,conn}   =
 \prod_{i=1}^{n} \left( - \frac{\partial}{\partial(a \zeta_{i})} \right)
  \frac{-1}{(n-2)!} \tilde{\Delta}_{n} |_{\Lambda_{i} = \Lambda} \;\;\;,
\eeqn
 where
\beqn
\label{eq:step2-2}
   \frac{1}{(n-2)!} \tilde{\Delta}_{n}      &\equiv&  \sum_{i_{1}}^{n}
\left(  \frac{\partial}{\partial \Lambda_{i_{1}} }\right)^{n-2}
   \int \cdots \int
 \sum_{\sigma \in {\cal S}_{n} }
 D_{n-2}( [ i_{1} - \sigma(i_{1})])
 \prod_{j (\neq i_{1})} D_{0}( [j- \sigma(j)])
  \nonumber \\
   &+&  \sum_{i_{1}, i_{2}}^{n}
\left(  \frac{\partial}{\partial \Lambda_{i_{1}} }\right)^{n-3}
\left(  \frac{\partial}{\partial \Lambda_{i_{2}} }\right)
   \int \cdots \int
 D_{n-3}( [ i_{1} - \sigma(i_{1})] )
  D_{1}( [ i_{2} - \sigma(i_{2})])
     \nonumber \\
 &\times& \prod_{j (\neq i_{1}, i_{2})} D_{0}( [j- \sigma(j)]) \nonumber \\
  &+& \cdots     \nonumber \\
  &+& \sum_{i_{1}, i_{2}, \cdots , i_{n-2}}^{n}
\left(  \frac{\partial}{\partial \Lambda_{i_{1}} }\right)
\left(  \frac{\partial}{\partial \Lambda_{i_{2}} }\right)
\cdots \left(  \frac{\partial}{\partial \Lambda_{i_{n-2}} }\right)
   \int \cdots \int  \nonumber \\
&~& \prod_{j=1}^{n-2} D_{1}( [ i_{j} - \sigma(i_{j})] )
 \prod_{j (\neq i_{1}, i_{2}, \cdots, i_{n-2})} D_{0}( [j- \sigma(j)])
   \;\;\;.
\eeqn

The integrals in the equation  above are  with respect to $z^{*}_{i}$'s and
 we  adopt a notation
\beqn
  [i] \equiv z_{i}^{*} \;\;\;, \;\;\;  [i-j] \equiv z_{i}^{*} - z_{j}^{*}
\;\;\;.
\eeqn

  This expansion is in one to one correspondence with the
   following expansion of
 $ \left( {\displaystyle \sum_{i=1}^{n}} x_{i} \right)^{n-2}$
\beqn
 &~&\left( \sum_{i=1}^{n} x_{i} \right)^{n-2}
 =  \sum_{i_{1}=1}^{n} x_{i_{1}}^{n-2}
 + ~_{n-2}C_{n-3}\left(\Theta(n \geq 5
 ) + \frac{1}{2} \delta_{n,4} \right) \sum_{i_{1}, i_{2}}x_{i_{1}}^{n-3}
 x_{i_{2}} \;\;\; \nonumber \\
 &+&~_{n-2}C_{n-4}\left(\Theta(n \geq 7
 ) + \frac{1}{2} \delta_{n,6} \right) \sum_{i_{1}, i_{2}}x_{i_{1}}^{n-4}
 x_{i_{2}}^{2}  \nonumber \\
 &+& ~_{n-2}C_{n-4}\left(\Theta(n \geq 6
 ) + \frac{1}{3} \delta_{n,5} \right)
 \sum_{i_{1}, i_{2}, i_{3} }x_{i_{1}}^{n-4}
 x_{i_{2}} x_{i_{3}} \;\;\; \nonumber \\
 &+&~_{n-2}C_{n-5}\left(\Theta(n \geq 9
  ) + \frac{1}{2} \delta_{n,8} \right) \sum_{i_{1}, i_{2}}x_{i_{1}}^{n-5}
 x_{i_{2}}^{3}   \;\;\; \nonumber \\
 &+&~_{n-2}C_{n-3} ~_{3}C_{2} \left(\Theta(n \geq 8
  ) + \frac{1}{2} \delta_{n,7} \right)
 \sum_{i_{1}, i_{2}, i_{3}}x_{i_{1}}^{n-5}
 x_{i_{2}}^{2} x_{i_{3}}   \;\;\; \nonumber \\
 &+&~_{n-2}C_{n-5}\left(\Theta(n \geq 9
 ) + \frac{1}{4} \delta_{n,6} \right) \sum_{i_{1}, i_{2},
 i_{3}, i_{4}  }x_{i_{1}}^{n-5}
 x_{i_{2}} x_{i_{3}} x_{i_{4}} \;\;\; \nonumber \\
 &+&  \cdots   \;\;\; \nonumber \\
 &+& \sum_{i_{1}, i_{2}, \cdots, i_{n-2} }x_{i_{1}} x_{i_{2}} \cdots
 x_{i_{n-2}} \;\;\;.
\eeqn
   Here the summations  without a parenthesis are over $k$ different
 integers $i_{1}, i_{2},  \cdots i_{k},$ $k= 1 \sim n-2$. The number
 of terms appearing is equal to the  number of
  partitions of $(n-2)$  objects into parts.

In order to put eqs.~(\ref{eq:step2-1}),  (\ref{eq:step2-2}) in
 a simpler  form, let us introduce
\beqn
\label{eq:braket}
  &&\left( \begin{array}{cccc}
 m_{1}, & m_{2}, & \cdots, & m_{n}  \\
 i_{1}, & i_{2}, &  \cdots, &  i_{n}
 \end{array} \right)_{n}  \nonumber \\
 &&\quad \equiv - \sum_{\sigma \in {\cal S}_{n} }
 D_{m_{1}}( [ i_{1} - \sigma(i_{1})])
  D_{m_{2}} ( [ i_{2} - \sigma(i_{2})] )
 \cdots  D_{m_{n}}  ( [ i_{k} - \sigma(i_{n})] )  \nonumber \\
&&\quad =-\sum_{\sigma\in{\cal S}_n } \frac1{[i_1-\sigma(i_1)]^{m_1+1} } \;
\frac1{[i_2-\sigma(i_2)]^{m_2+1} } \;\cdots \frac1{[i_n-\sigma(i_n)]^{m_n+1} }
\;\;.\eeqn
In particular,
\beqn
   \left( \begin{array}{cccc}
 n-2,   & 0,       & \cdots,       &  0 \\
 i_{1}, & i_{2}, & \cdots, & 0
 \end{array} \right)_{n}
  &\equiv& - \sum_{\sigma \in {\cal S}_{n} }
 D_{n-2}( [ i_{1} - \sigma(i_{1})])
 \prod_{j (\neq i_{1})} D_{0}( [j- \sigma(j)])  \nonumber \\
  \left( \begin{array}{ccccc}
 n-3,   & 1,       & 0,        & \cdots,   &  0 \\
 i_{1}, & i_{2}, &  i_{3}, &  \cdots,  &  0
 \end{array} \right)_{n}
  &\equiv&    - \sum_{\sigma \in {\cal S}_{n} }
  D_{n-3}( [ i_{1} - \sigma(i_{1})] )
  D_{1}( [ i_{2} - \sigma(i_{2})])  \nonumber \\
  &~& \times \prod_{j (\neq i_{1}, i_{2})} D_{0}( [j- \sigma(j)]) \nonumber \\
  &{\rm e.t.c}&
\eeqn

  In the appendix A, we prove  that
\beqn
\label{eq:sub-recursion}
  \left( \begin{array}{cccc}
 m_{1}, & m_{2},  & \cdots, & m_{n} \\
 i_{1},   & i_{2},    &  \cdots, & i_{n}
 \end{array} \right)_{n}   = 0 \;\;\; {\rm if} \;\;\ \sum_{\ell} m_{\ell}
 \leq  n-3 \;\;\;  ,
\eeqn
 as well as
\beqn
\label{eq:recursion}
 &~& \left( \begin{array}{ccccccc}
 m_{1}, & m_{2}, & \cdots, & m_{k}, & 0, & \cdots,  & 0 \\
 i_{1},   & i_{2},   & \cdots, &   i_{k},& i_{k+1}, & \cdots,  & 0
 \end{array} \right)_{n}     \nonumber \\
 &=&  \sum_{\ell= 1}^{k} \frac{1}{\left[ i_{\ell}- i_{n} \right]^{2}}
  \left( \begin{array}{clccccr}
 m_{1}, & \cdots,   &  m_{\ell}-1, & \cdots,  & m_{k}, & 0,  & \cdots \\
 i_{1}, & \cdots,    &  i_{\ell},  &  \cdots, &  i_{k},& \cdots, &\cdots
 \end{array} \right)_{n-1}      \;\;\; \nonumber \\
 &{\rm if }& \;\;\; \sum_{\ell} m_{\ell} = n-2 \;\;\;.
\eeqn
 In particular,
\beqn
\label{eq:n-2,0}
   \left( \begin{array}{lcr}
 n-2, & 0, & \cdots  \\
 i_{1}, &\cdots, & \cdots
 \end{array} \right)_{n}
=  \frac{1}{\left[i_{1} - i_{n}\right]^{2} }
   \left( \begin{array}{lcr}
 n-3, & 0, & \cdots  \\
 i_{1}, & \cdots, & \cdots
 \end{array} \right)_{n-1}   = \frac{1}{ {\displaystyle
 \prod_{j (\neq i_{1})}^{n}  }
\left[ i_{1} -j \right]^{2} } \;\;\;
\eeqn
  and
\beqn
\label{eq:n-3,1}
 \left( \begin{array}{lcr}
 n-3, & 1, & \cdots  \\
 i_{1}, & i_{2} & \cdots
 \end{array} \right)_{n}
 &=&
  \frac{1}{\left[ i_{1} - i_{n}\right]^{2} } \left( \begin{array}{lcr}
 n-4, & 1, & \cdots  \\
 i_{1}, & i_{2}, & \cdots
 \end{array} \right)_{n-1}  \nonumber \\
   ~&+&
  \frac{1}{\left[ i_{2} - i_{n}\right]^{2} } \left( \begin{array}{clcr}
 n-3, & 0, & \cdots  \\
 i_{1}, & i_{2}, & \cdots
 \end{array} \right)_{n-1}   \;\;\;.
\eeqn

Let us introduce graphs in which    the factor $1/[i-j]^2$  is represented
 by  a double line linking  circle $i$ and circle $j$
  to handle the  quantities defined by eq.~(\ref{eq:braket})
 more easily.
For example, for n=3,
\beq
 \left( \begin{array}{ccc}
 1 & 0 & 0  \\
 1 & 2 & 3
 \end{array} \right)_{3}
 =\frac1{[z^*_1-z^*_2]^2}\; \frac1{[z^*_1-z^*_3]^2}
 \equiv \epscenterboxy {2.2cm}{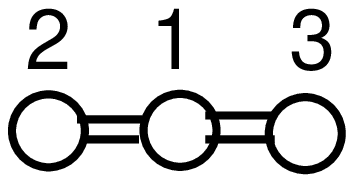}
  .
\eeq
Using the recursin relation eq.~(\ref{eq:recursion})
 and eq.~(\ref{eq:sub-recursion}) we have, for n=4,
\beqn
 &&\left( \begin{array}{cccc}
 2 & 0 & 0 & 0  \\
 1 & 2 & 3 & 4
 \end{array} \right)_{4}
 = \frac{1}{[z_1^*-z_4^*]^2}
 \left( \begin{array}{ccc}
 2-1 & 0 & 0   \\
 1 & 2 & 3
 \end{array} \right)_{3} \nonumber \\
 &&\qquad\qquad = \epscenterboxy{2.5cm}{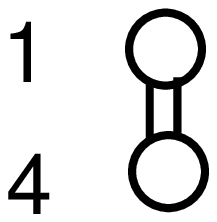}\times
 \epscenterboxy{2.2cm}{ai_fig/lk213.eps}
 =\epscenterboxy{2.5cm}{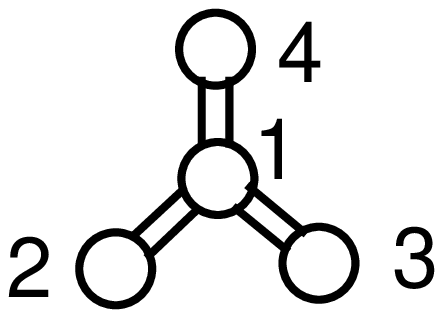} ,
\eeqn
\beqn
  &&\left( \begin{array}{cccc}
 1 & 1 & 0 & 0  \\
 1 & 2 & 3 & 4
 \end{array} \right)_{4}
 =\frac{1}{[z_1^*-z_4^*]^2}
 \left( \begin{array}{ccc}
 1-1 & 1 & 0   \\
 1 & 2 & 3
 \end{array} \right)_{3}
 + \frac{1}{[z_2^*-z_4^*]^2}
 \left( \begin{array}{ccc}
 1 & 1-1 & 0   \\
 1 & 2 & 3
 \end{array} \right)_{3} \nonumber \\
 &&\qquad\qquad = \epscenterboxy{2.5cm}{ai_fig/lk14.eps}\times
 \epscenterboxy{2.2cm}{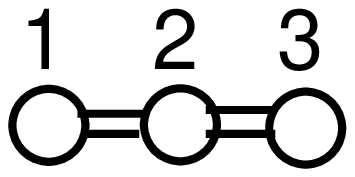}
 + \epscenterboxy{2.5cm}{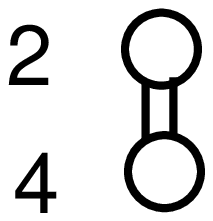}\times
 \epscenterboxy{2.2cm}{ai_fig/lk213.eps} \nonumber \\
 &&\qquad\qquad =  \epscenterboxy{2.cm}{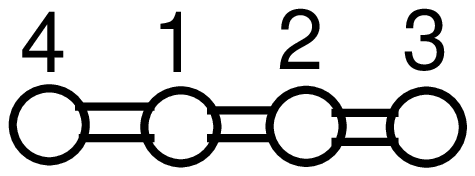}
 + \epscenterboxy{2.cm}{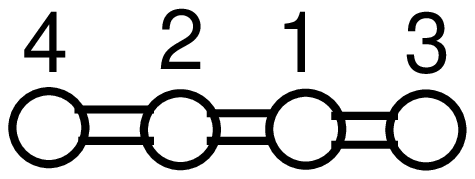}  ,
\eeqn
  and, for n=5,
\beqn
 &&\left( \begin{array}{ccccc}
 3 & 0 & 0 & 0 & 0  \\
 1 & 2 & 3 & 4 & 5
 \end{array} \right)_{5}
 =\frac{1}{[z_1^*-z_5^*]^2}
 \left( \begin{array}{cccc}
 3-1 & 0 & 0 & 0   \\
 1 & 2 & 3 & 4
 \end{array} \right)_{4} \nonumber \\
 &&\qquad\qquad = \epscenterboxy{2.5cm}{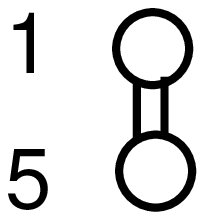}\times
 \epscenterboxy{2.5cm}{ai_fig/lk1_234.eps}
 =\epscenterboxy{2.5cm}{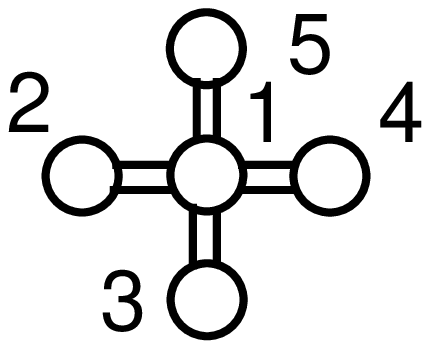} ,
\eeqn
\beqn
 &&\left( \begin{array}{ccccc}
 2 & 1 & 0 & 0 & 0  \\
 1 & 2 & 3 & 4 & 5
 \end{array} \right)_{5}
 =\frac{1}{[z_1^*-z_5^*]^2}
 \left( \begin{array}{cccc}
 2-1 & 1 & 0 & 0   \\
 1 & 2 & 3 & 4
 \end{array} \right)_{4}
 + \frac{1}{[z_2^*-z_5^*]^2}
 \left( \begin{array}{cccc}
 2 & 1-1 & 0 & 0   \\
 1 & 2 & 3 & 4
 \end{array} \right)_{4} \nonumber \\
 &&\qquad\qquad =  \epscenterboxy{2.5cm}{ai_fig/lk15.eps}\times
 \Biggl\{ \epscenterboxy{2.cm}{ai_fig/lk4123.eps}
 +\epscenterboxy{2.cm}{ai_fig/lk4213.eps}
 \Biggr\}  \nonumber \\
 &&\qquad\qquad \qquad+ \epscenterboxy{2.5cm}{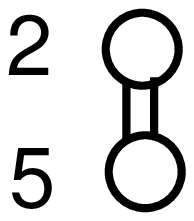}\times
 \epscenterboxy{2.5cm}{ai_fig/lk1_234.eps} \nonumber \\
 &&\qquad\qquad =  \epscenterboxy{2.5cm}{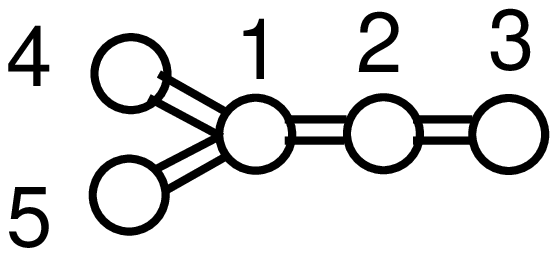}
 + \epscenterboxy{2.5cm}{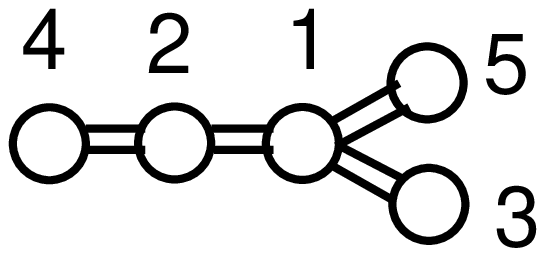}
+ \epscenterboxy{2.5cm}{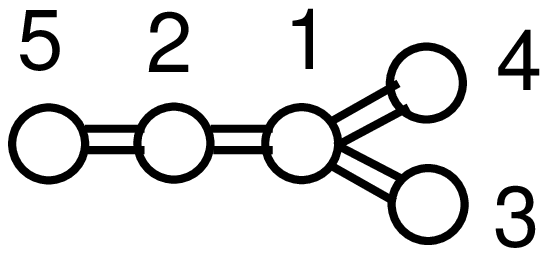}  .
\eeqn
 From the examples  above, it is clear that the graphs for the general
 case can be written down quite easily.

In terms of the quantities defined by eq.~(\ref{eq:braket}), we obtain
 a formula for the n-point resolvent :
\beqn
\label{eq:result}
 \left(\frac{N}{\Lambda}\right)^{n-2} << \prod_{i=1}^{n} tr
 \frac{1}{p_{i} - \hat{M}} >>_{N,conn}   =
 \prod_{i=1}^{n} \left( - \frac{\partial}{\partial(a \zeta_{i})} \right)
  \frac{-1}{(n-2)!} \tilde{\Delta}_{n} |_{\Lambda_{i} = \Lambda} \;\;\;,
\eeqn
where
\beqn
\label{eq:Delta}
 &&-\frac1{(n-2)!} \tilde{\Delta}_n(z^*_1,\cdots.z^*_n)  \nonumber \\
 &&\qquad\qquad
 = \sum_{i_1}^{n} \Bigl(\frac{\p}{\p\Lambda_{i_1} }\Bigr)^{n-2}
 \int \cdots \int \;
 \left( \begin{array}{cccc}
 n-2, & 0, & \cdots, & 0 \\
 i_1 , & i_2, & \cdots, & i_n
 \end{array} \right)_n   \nonumber \\
 &&\qquad
 + \sum_{m_1+m_2=n-2} \sum_{ (i_1,i_2) }
 \Bigl(\frac{\p}{\p\Lambda_{i_1} }\Bigr)^{m_1}
 \Bigl(\frac{\p}{\p\Lambda_{i_2} }\Bigr)^{m_2}
 \int \cdots \int \;
 \left( \begin{array}{ccccc}
 m_1, & m_2,& 0, & \cdots, & 0 \\
 i_1 , & i_2, & i_3, & \cdots, & i_n
 \end{array} \right)_n   \nonumber \\
 &&\qquad
 + \sum_{ {m_1+m_2+m_3\atop =n-2 }} \sum_{ (i_1,i_2,i_3) }
 \Bigl(\frac{\p}{\p\Lambda_{i_1} }\Bigr)^{m_1}
 \Bigl(\frac{\p}{\p\Lambda_{i_2} }\Bigr)^{m_2}
 \Bigl(\frac{\p}{\p\Lambda_{i_3} }\Bigr)^{m_3}
 \int \cdots \int \;
 \left( \begin{array}{cccccc}
 m_1, & m_2, & m_3,& 0, & \cdots, & 0 \\
 i_1 , & i_2, & i_3, & i_4, & \cdots, & i_n
 \end{array} \right)_n   \nonumber \\
 &&\qquad +\qquad \cdots \nonumber \\
&&\qquad
 + \sum_{ (i_1,\cdots,i_{n-2}) }
 \Bigl(\frac{\p}{\p\Lambda_{i_1} } \Bigl)
 \cdots
 \Bigl(\frac{\p}{\p\Lambda_{i_{n-2}} } \Bigl)
 \int \cdots \int \;
 \left( \begin{array}{ccccc}
 1, & \cdots, & 1,& 0,  & 0 \\
 i_1 , & \cdots, & i_{n-2}, &  i_{n-1}, & i_n
 \end{array} \right)_n
\quad .
\eeqn
  Here $m_{\ell} \geq 1$  and  the summation $(i_{1}, \cdots,  i_{k})$
  denotes  a set of  $k$ unequal integers from $1,2, \cdots, n$  and
 $i_{k+1}, \cdots, i_{n}$
  in the array represents  the remaining integers.
Eqs.~(\ref{eq:result}), (\ref{eq:Delta}) are  a part of our main results.

  It is straightforward to  write down  the correlator for lower  $n$
 explicitly;
\beqn
   \frac{\partial}{\partial \Lambda}
 \frac{-1}{(0)!} \tilde{\Delta}_{2}    &=&  \sum_{i=1}^{2}
 \frac{\p z^*_i}{\p \Lambda_i}
 \frac{1}{ {\displaystyle
 \prod_{j (\neq i)}^{2} } [i-j] } \;\;,   \nonumber \\
 \frac{-1}{(1)!} \tilde{\Delta}_{3}   &=&
 \sum_{i=1}^{3}
 \frac{\p z^*_i}{\p \Lambda_i}
 \frac{1}{ {\displaystyle \prod_{j (\neq i)}^{3} }
 [i-j] } \;\;\; .
\eeqn
For $n=4, 5$, the correlator   can be written more compactly
  using graphs as  introduced below:
\beqn
\label{eq:n=4}
  \frac{-1}{2!} \tilde{\Delta}_{4}
 &=&
 \sum \; \frac{ \p}{\p\Lambda_{i_1} }
 \Biggl\{ \epscenterboxy{2.5cm}{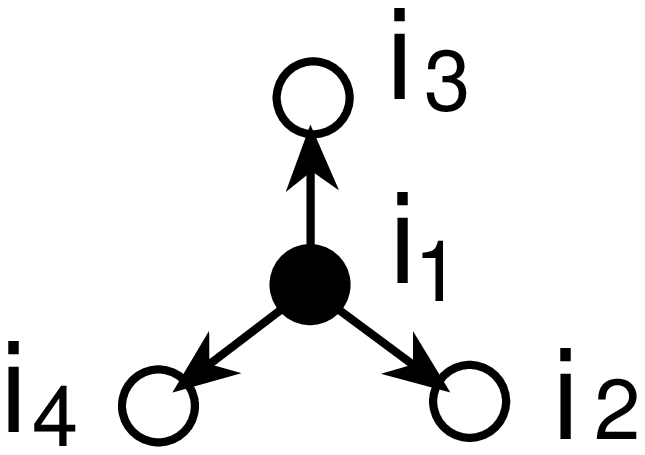} \Biggr\} \nonumber \\
 &+&
 \sum \; \Biggl\{ \epscenterboxy{1.8cm}{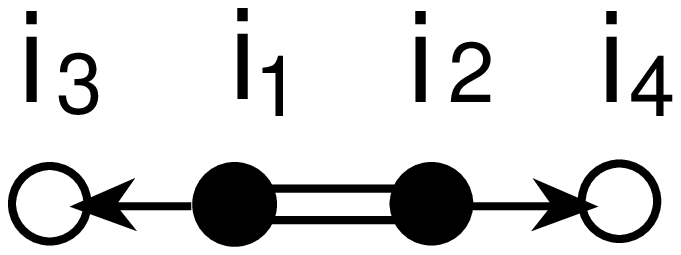} \Biggr\}
 \quad  ,
\eeqn
\beqn
\label{eq:n=5}
 \frac{-1}{3!} \tilde{\Delta}_{5}
 &=&
 \sum \; \Bigl(\frac{ \p}{\p\Lambda_{i_1} }\Bigr)^2
 \Biggl\{ \epscenterboxy{2.7cm}{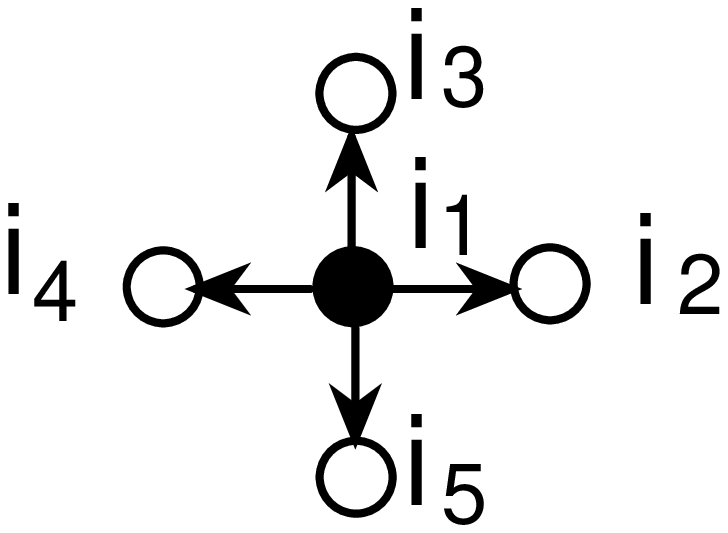} \Biggr\} \nonumber \\
 &+&
  \sum \; \Bigl(\frac{ \p}{\p\Lambda_{i_1} }\Bigr)
 \Biggl\{ \epscenterboxy{2.5cm}{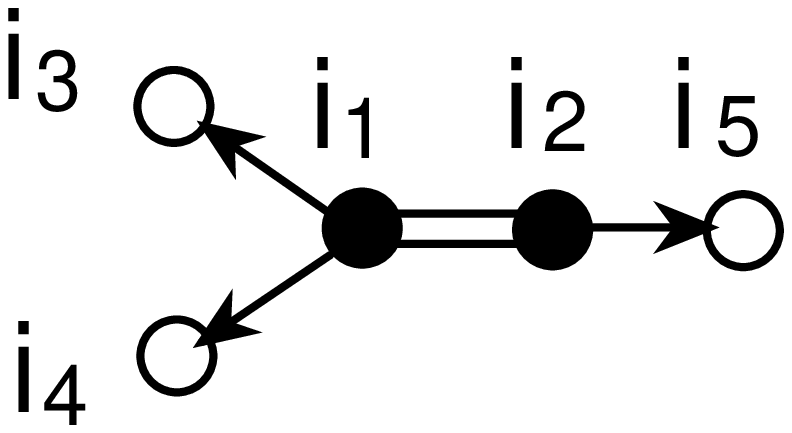} \Biggr\} \nonumber \\
 &+&
 \sum \; \Biggl\{ \epscenterboxy{1.8cm}{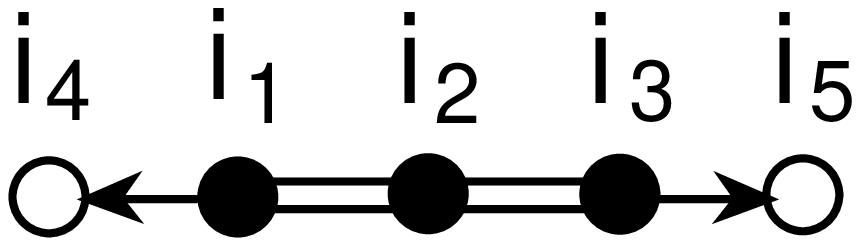} \Biggr\}
 \quad.
\eeqn
In these figures a double line linking circle $i$ and circle $j$,
  a single line
  having an  arrow from circle $i$ to circle $j$  and a solid circle $i$
 represent
 $1/[i-j]^2$, $1/[i-j]$ and  $\frac{\p z^*_i}{\p \Lambda_i}$ respectively.
The summations are over all possible graphs that
 have the same topology specified.
Each graph appears just for once in the summation.
Note that the links  to the external circles
 are not double lines but   the single ones with arrows and that the
 internal circles are solid circles.

For general $n$, $-\frac{1}{(n-2)!}\tilde{\Delta}_n$ is expressed in
 the same way.
The rule is as follows.
First, we consider all possible graphs which have $n$ circles and
$n-1$ links in the same way as
  in the case $n=5$.
Second, if  the internal solid circle $i$ has  $\ell_{i}$ links in
 each graph, the graph is operated by
$\prod_i \Bigl (\frac{\p}{\p \Lambda_i} \Bigl)^{\ell_{i-2} }$.
Then the summation over  all graphs gives the expression
 for  $-\frac{1}{(n-2)!}\tilde{\Delta}_n$.

 Isolating the term which comes with the highest number of
 total derivatives
  of the bare cosmological constant $\Lambda$ ( as opposed to
  $\Lambda_{i}$)
  by
\beqn
  \sum_{i_{1}} \left( \frac{\partial}{\partial \Lambda_{i_{1}}}\right)^{n-2}
   &=&  \left( \frac{\partial}{\partial \Lambda} \right)^{n-3}
  \sum_{i_{1} =1}^{n} \frac{\partial}{\partial \Lambda_{i_{1}} }
    \nonumber \\
 &-& \sum_{i_{1}=1}^{n} \left[ \sum_{k=0}^{n-4}
 \left( \frac{\partial}{\partial \Lambda_{i_{1}} } \right)^{n-k-3}
  \left(  \sum_{\ell=1}^{n} \frac{\partial}{\partial \Lambda_{\ell} }
 \right)^{k} \right]  \sum_{j (\neq i_{1})}  \frac{\partial}{\partial
  \Lambda_{j}} \;\;\;,   \nonumber
\eeqn
 we obtain another expression for $\frac{-1}{(n-2)!}\tilde{\Delta}_n$ :
\beqn
\label{eq:deformed-Delta}
 &~&  \frac{-1}{(n-2)!} \tilde{\Delta}_{n}  =
 \left( \frac{\partial}{\partial \Lambda} \right)^{n-3}
   \sum_{i_{1}}^{n}
\left(  \frac{\partial}{\partial \Lambda_{i_{1}} }\right)
   \int \cdots \int
  \left( \begin{array}{clcr}
 n-2, & 0, & \cdots & ~ \\
 i_{1}, & \cdots, & \cdots & ~
 \end{array} \right)   \nonumber \\
 &+&  \sum_{i_{1}, i_{2}}^{n}
\left(  \frac{\partial}{\partial \Lambda_{i_{1}} }\right)^{n-3}
  \left(  \frac{\partial}{\partial \Lambda_{i_{2}} }\right)
   \int \cdots \int \left\{
 \left(\Theta(n \geq 5
 ) + \frac{1}{2} \delta_{n,4} \right)
    \left( \begin{array}{clcr}
 n-3, & 1, & 0, &  \cdots \\
 i_{1}, & i_{2}, & \cdots,  & \cdots
 \end{array} \right)  \right.   \nonumber \\
 &~& \left.  -(n-3)    \left( \begin{array}{clcr}
 n-2, & 0, & \cdots & ~ \\
 i_{1}, & i_{2},& \cdots  & ~
 \end{array} \right)   \right\}      \;\;\; \nonumber \\
   &+&  \sum_{i_{1}, i_{2}}^{n}
\left(  \frac{\partial}{\partial \Lambda_{i_{1}} }\right)^{n-4}
\left(  \frac{\partial}{\partial \Lambda_{i_{2}} }\right)^{2}
   \int \cdots \int
  \left\{ \left(\Theta(n \geq 7
 ) + \frac{1}{2} \delta_{n,6} \right)   \left( \begin{array}{clcr}
 n-4, & 2, & 0 & \cdots \\
 i_{1}, & i_{2}, & \cdots & \cdots
 \end{array} \right)   \right.  \nonumber \\
 &~&  \left.  - \left(\sum_{k=0}^{n-4}~_{k}C_{1} \right)
   \left( \begin{array}{clcr}
 n-2, & 0, & \cdots & ~ \\
 i_{1}, & i_{2}, & \cdots & ~
 \end{array} \right) \right\}
   \;\;\; \nonumber \\
    &+&  \sum_{i_{1}, (i_{2}, i_{3})}^{n}
\left(  \frac{\partial}{\partial \Lambda_{i_{1}} }\right)^{n-4}
\left(  \frac{\partial}{\partial \Lambda_{i_{2}} }\right)
\left(  \frac{\partial}{\partial \Lambda_{i_{3}} }\right)
   \int \cdots \int \left\{
\left(\Theta(n \geq 6
 ) + \frac{1}{3} \delta_{n,5} \right)  \times
        \right.  \nonumber \\
 &~&   \left( \begin{array}{clcr}
 n-4, & 1, & 1 & 0 \\
 i_{1}, & i_{2} & i_{3} & ~
 \end{array} \right)
  \left. -   \left( \sum_{k=0}^{n-4} ~_{k}C_{1} \right)
   \left( \begin{array}{clcr}
 n-2, & 0, & 0, & \cdots \\
 i_{1}, & i_{2} & i_{3}, & \cdots
 \end{array} \right) \right\}
 \;\;\; \nonumber \\
   &+&  \sum_{i_{1}, i_{2}}^{n}
\left(  \frac{\partial}{\partial \Lambda_{i_{1}} }\right)^{n-5}
\left(  \frac{\partial}{\partial \Lambda_{i_{2}} }\right)^{3}
   \int \cdots \int \left\{
\left(\Theta(n \geq 9
  ) + \frac{1}{2} \delta_{n,8} \right)
 \left( \begin{array}{clcr}
 n-5, & 3, & \cdots & ~ \\
 i_{1}, & i_{2} & ~ & ~
 \end{array} \right)  \right.  \nonumber \\
  &~&  \left.
  -  \left( \sum_{k=0}^{n-4} ~_{k}C_{2} \right)
 \left( \begin{array}{clcr}
 n-2, & 0, & \cdots & \cdots \\
 i_{1}, & i_{2}, & \cdots & \cdots
 \end{array} \right) \right\}
   \;\;\; \nonumber \\
    &+&  \sum_{i_{1}, i_{2}, i_{3}}^{n}
\left(  \frac{\partial}{\partial \Lambda_{i_{1}} }\right)^{n-5}
\left(  \frac{\partial}{\partial \Lambda_{i_{2}} }\right)^{2}
\left(  \frac{\partial}{\partial \Lambda_{i_{3}} }\right)
   \int \cdots \int  \left\{
\left(\Theta(n \geq 8
 ) + \frac{1}{2} \delta_{n,7} \right) \times  \right.  \nonumber \\
 &~& \left.
  \left( \begin{array}{clcr}
 n-5, & 2, & 1, & \cdots \\
 i_{1}, & i_{2}, & i_{3}, & \cdots
 \end{array} \right)
  - \left( \sum_{k=0}^{n-4} ~_{k}C_{2}  \right)
  \left( \begin{array}{clcr}
 n-2, & 0, & 0 & \cdots \\
 i_{1}, & i_{2}, &i_{3},  & \cdots
 \end{array} \right)   \right\}
          \;\;\;   \\
    &+&  \sum_{i_{1}, (i_{2}, i_{3}, i_{4}) }^{n}
\left(  \frac{\partial}{\partial \Lambda_{i_{1}} }\right)^{n-5}
\left(  \frac{\partial}{\partial \Lambda_{i_{2}} }\right)
\left(  \frac{\partial}{\partial \Lambda_{i_{3}} }\right)
\left(  \frac{\partial}{\partial \Lambda_{i_{4}} }\right)
   \int \cdots \int
   \left\{ \left(\Theta(n \geq 7
 ) + \frac{1}{4} \delta_{n,6} \right) \times   \right.   \nonumber \\
  &~&  \left.
   \left( \begin{array}{clccr}
 n-3, & 1, & 1, & 1, & \cdots \\
 i_{1}, & i_{2}, & i_{3}, & i_{4}, &\cdots
 \end{array} \right)
   -  \left(\sum_{k=0}^{n-4} ~_{k}C_{2} \right)
  \left( \begin{array}{clccr}
 n-2, & 0, & 0, & 0, &\cdots \\
 i_{1}, & i_{2}, & i_{3}, & \cdots
 \end{array} \right) \right\}
     \;\;\; \nonumber \\
  &+& \cdots     \nonumber \\
 &+& \sum_{ (i_{1}, i_{2}, \cdots , i_{n-2}) }^{n}
\left(  \frac{\partial}{\partial \Lambda_{i_{1}} }\right)
\cdots \left(  \frac{\partial}{\partial \Lambda_{i_{n-2}} }\right)
   \int \cdots \int
  \left\{  \left( \begin{array}{clclcr}
 1, & 1, & \cdots & 1, & 0, & \cdots \\
 i_{1}, & i_{2}, & \cdots & i_{n-2}, &\cdots &\cdots
 \end{array} \right)  \right. \nonumber \\
 &-&  \left.  \left( \begin{array}{clclcr}
 n-2, & 0, & \cdots & 0, & 0, & 0 \\
 i_{1}, & i_{2}, & \cdots & i_{n-2}, &\cdots &\cdots
 \end{array} \right)
- ....  -
  \left( \begin{array}{clclcr}
 0, & 0, & \cdots & n-2, & 0& 0 \\
 i_{1}, & i_{2}, & \cdots & i_{n-2}, &\cdots &\cdots
 \end{array} \right)
       \right\} \;.  \nonumber
\eeqn
   Here   the summations with a parenthesis $(i_{1}, i_{2}, \cdots, i_{k})$
   are over $k$ unequal indistinguished indices.
  Using this formula  and the partial fraction, one can put  the expression
  into a form which does not contain  a direct link among solid circles.
   This is sometimes useful in the manipulation we carry out later.

   The following parametrization of  $z_{i}^{*}$ \cite{DKK} is important
  in the next section:
\beq
\label{eq:z=}
  z_{i}^{*} = \exp (2\eta \cosh \theta_{i}) \;\;\;.
\eeq
    This parametrization is understood  together with
\beq
 p_{i}- p_{i}^{*}=
  [\hat{M}](z_{i}; \Lambda_{i}) -[\hat{M}]^{*} =
 a M \cosh m\theta_i \quad,\quad
 \eta=(a M/2)^{1/m}
\label{3loop-o}
\eeq
and
\beq
 \Lm-\Lm_*=-(m+1) \eta^{2m}=-(m+1)a^2 t\;, \;\; M = (2t)^{1/2}\;\;\;.
\label{3loop-p}
\eeq
  We  obtain
\beqn
  \frac{\partial \theta_{i}}{ \partial \Lambda} \mid_{\zeta_{i}}
  = -\frac{1}{\eta} \frac{\partial \eta}{\partial \Lambda}
  \frac{\cosh m\theta_{i}}{ \sinh m \theta_{i}} \;\;\;,
 \;\;\;\;
 \frac{\p z^*_i}{\p \Lambda_i} = 2 \left(
 \frac{\partial \eta}{\partial \Lambda} \right)
  \frac{ \sinh (m-1) \theta_{i} }{ \sinh m \theta_{i} }\;\;\;.
\eeqn
   The origin of  the parametrization  eq.~(\ref{eq:z=})   comes from
   the planar solution to the Heisenberg algebra \cite{Doug}  in \cite{DKK}.
   In fact, eq.~(\ref{3loop-o})  represents  the solution
  at the $(m+1,m)$ critical point.
( $ p_{i}^{*}$ and $[\hat{M}]^{*} $ represent  the critical value to
$p_{i}$  and  to $[\hat{M}]$  respectively.)


\section{ Fusion Rules,  Crossing Symmetry and  Polygons Associated}

  We now discuss
  the term we have isolated in  eq.~(\ref{eq:deformed-Delta}),
  the expression  for  the $n$-point
  resolvent,  namely the one which comes with
  the highest number of total derivatives with respect  to $\Lambda$
   and  is, therefore, familiar from the case of pure two-dimensional gravity.
    We will exhibit striking properties  with this term,
  using the parametrization noted in eq.~(\ref{eq:z=}).
   Let us  define
\beqn
\label{eq:polygon}
    P_{n}(\theta_{1}, \theta_{2}, \cdots \theta_{n})
  &\equiv&
   \sum_{i_{1}}^{n}
\left(  \frac{\partial}{\partial \Lambda_{i_{1}} }\right)
   \int \cdots \int
  \left( \begin{array}{clcr}
 n-2, & 0, & \cdots, & 0 \\
 i_{1}, &  i_{2}, &  \cdots,  &  i_{n}
 \end{array} \right)
 \nonumber \\
 &=&  \sum_{i=1}^{n}
\frac{\p z^*_i}{\p \Lambda_i}
 \frac{1} { {\displaystyle
 \prod_{j (\neq i)}^{n} } [i-j] } \;\;\;.
\eeqn
    A key manipulation we  will use is   the  partial fraction
\beq
\label{eq:pf}
  \frac{1}{[i-j][i-k]} = \frac{1}{[i-k][k-j]} + \frac{1}{[i-j][j-k]} \;\;\;.
\eeq
  One can associate a line from $i$ to $j$ with $\frac{1}{[i-j]}$.
 The following identity
  is responsible for expressing  $P_{n}$ as
 a sum of the product of  $n$ factors each of which depends
  only on  $\theta_{i}$:
\beqn
   I_{k} ( \alpha, \beta ;m ) &\equiv&
 \lefteqn{
 \frac{1}{\cosh\alpha-\cosh\beta}
 \left(\;\frac{\sinh (m-k)\alpha}{\sinh m\alpha}
 -\frac{\sinh (m-k)\beta}{\sinh m\beta} \;\right) \mbox{                 }
 } \nonumber\\
 &&=-2\sum_{j=1}^{m-k}\sum_{i=1}^{k}
 \;\frac{\sinh (m-j-i+1)\alpha}{\sinh m\alpha}
 \;\frac{\sinh (m-j-k+i)\beta}{\sinh m\beta}
 \quad .
\label{eq:ident}
\eeqn

Let us first work out  the cases $n=2,3,4$  to get a feeling.  For $n=2$,
\beqn
 &~&  P_{2}(\theta_{1}, \theta_{2})
 =  \frac{\frac{\p z^*_1}{\p \Lambda_1} -\frac{\p z^*_2}{\p \Lambda_2}}
 {[1-2]} = 2 \left(
 \frac{ \partial \eta}{\partial \Lambda} \right) \frac{1}{2\eta}  I_{k_{1}= 1}
 \left( \theta_{1}, \theta_{2} ; m \right)  \nonumber \\
 &=& 2 \left( \frac{ \partial \eta}{\partial \Lambda}\right)
\left( \frac{-1}{\eta} \right) \sum_{j=1}^{m-k_{1}}\sum_{i_{1}=1}^{k_{1}}
\frac{\sinh (m-j_{1}-i_{1}+1)\theta_{1} }{\sinh m\theta_{1}}
 \frac{\sinh (m-j_{1}+i_{1}- k_{1})\theta_{2}}{\sinh m\theta_{2}}
 \nonumber \\
 &=& 2 \left( \frac{ \partial \eta}{\partial \Lambda} \right)
\left(\frac{-1}{\eta} \right)
\sum_{j_{1}=1}^{m-1}
 \;\frac{\sinh (m-j_{1})\theta_{1} }{\sinh m\theta_{1}}
 \;\frac{\sinh (m-j_{1})\theta_{2}}{\sinh m\theta_{2}} \;\;,
\eeqn
   where $i_{1}=k_{1}=1$.
  For $n=3$, we use eq.~(\ref{eq:pf}) for the  term
 containing $\frac{\p z^*_2}{\p \Lambda_2}$ to create a  link
  $[1-3]$, which is originally absent.
 This relates $P_{3}$ to $P_{2}$.    We find
\beqn
&~& P_{3} (\theta_{1}, \theta_{2},\theta_{3})  \nonumber \\
&=&  2 \left( \frac{ \partial \eta}{\partial \Lambda} \right)
 \frac{ I_{1}\left( \theta_{2}, \theta_{1} ; m \right)-
 I_{1}\left( \theta_{2}, \theta_{3} ; m \right) }  {[1-3]}
  \nonumber \\
 &=& 2 \left( \frac{ \partial \eta}{\partial \Lambda}\right)
 \left(\frac{-1}{\eta}\right)^{2}
 \left(\sum_{j_{1}=1}^{m-k_{1}}\sum_{i_{1}=1}^{k_{1}=1} \right)
 \left(\sum_{j_{2}=1}^{m-k_{2}}\sum_{i_{2}=1}^{k_{2}} \right)  \\
&~& \frac{\sinh (m-j_{2}-i_{2}+1)\theta_{1} }{\sinh m\theta_{1}}
\frac{\sinh (m-j_{1}+i_{1}- k_{1})\theta_{2}}{\sinh m\theta_{2}}
 \frac{\sinh (m-j_{2}+i_{2}- k_{2})\theta_{3}}{\sinh m\theta_{3}}
  \;. \nonumber
\eeqn
   Here $k_{2} = j_{1} + i_{1}-1= j_{1}$.

This can be repeated  for arbitrary $n$.
 In the case $n=4$,  we use the partial fraction
 for  the two terms containing $\frac{\p z^*_2}{\p \Lambda_2}$ and
 $\frac{\p z^*_3}{\p \Lambda_3}$
  to create a link $[1-4]$, which is originally absent.
 This enables us to relate the  case $n=4$
  to the case $n=3$.
  In general,  $P_{n}$  is  related to $P_{n-1}$
by using  the partial fraction for the terms containing
 $\frac{\p z^*_2}{\p \Lambda_2} \sim \frac{\p z^*_{n-1}}{\p \Lambda_{n-1}}$
   to create a link $[1-n]$.
   We obtain
\beqn
\label{eq:answer}
 &~& P_{n}(\theta_{1}, \theta_{2}, \cdots \theta_{n})  =
  - \left( \frac{ \partial \eta^{2}}{\partial \Lambda}\right)
 \left(\frac{-1}{\eta}\right)^{n}
 \left( \prod_{\ell=1}^{n-1}
\sum_{j_{\ell}=1}^{m-k_{\ell}}\sum_{i_{\ell}=1}^{k_{\ell}} \right)
\nonumber \\
 &~& \left( \prod_{\ell^{\prime}=1}^{n-1}
 \;\frac{\sinh (m-j_{\ell^{\prime}}+
i_{\ell^{\prime}}- k_{\ell^{\prime} })\theta_{\ell^{\prime}+1 }
 }{\sinh m\theta_{\ell^{\prime}+1 }  }  \right)
 \frac{\sinh (m-j_{n-1}-i_{n-1}+1)\theta_{1} }{\sinh m\theta_{1}} \;\;\;,
\eeqn
  where $k_{\ell} = j_{\ell-1} + i_{\ell-1} -1,$~~ for
$\ell = 2,3, \cdots , (n-1) $.
	Eq.~(\ref{eq:answer})  expresses
 the $P_{n}(\theta_{1}, \theta_{2}, \cdots \theta_{n})$
  as a sum of the $n$-products
  of  the factor ${\displaystyle  \frac{\sinh(m-k)\theta_{i}}
{\sinh m\theta_{i}} }$.
 Owing to this property,  one can perform
 the inverse Laplace transform immediately, which we will  carry out at
 eq.~(\ref{eq:fusionanswer}).

  Let us now  discuss  the restrictions on the summations of
 $2n-3$  integers $j_{1},$ $i_{2}, j_{2},$ $\cdots$
  $i_{n-1}, j_{n-1}$ in  eq.~(\ref{eq:answer}).
  We   write these as  a set:
\beqn
 &~&{\cal F}_{n}( j_{1}, i_{2}, j_{2} \cdots
  i_{n-1}, j_{n-1})   \nonumber \\
 &\equiv& \{ ( j_{1}, i_{2}, j_{2}, \cdots
  i_{n-1}, j_{n-1}) \mid
 1 \leq i_{\ell} \leq k_{\ell}, \;
 1 \leq j_{\ell} \leq m- k_{\ell}, \;  {\rm for}~ \ell =
  1,2, \cdots n-1 \}  \nonumber \\
 &=& {\cal F}_{2}( i_{1}=1, j_{1};k_{1}=1)
 \prod_{\ell=2}^{n-1} \cap {\cal F}_{2}( i_{\ell}, j_{\ell};k_{\ell})
 \;,
\eeqn
  where
\beqn
  {\cal F}_{2}( i_{\ell}, j_{\ell};k_{\ell})
 \equiv  \{ (  i_{\ell}, j_{\ell} ) \mid
 1 \leq i_{\ell} \leq k_{\ell},\;
 1 \leq j_{\ell} \leq m- k_{\ell},\;\;{\rm with}\; k_{\ell}~~{\rm fixed} \}
 \;\;.
\eeqn
  We will show that  these restrictions on the sums
  are in fact in one-to-one correspondence with
 the fusion rules of the unitary minimal models
 for the diagonal primaries.
  Let us begin with the case $n=3$.
  Define
\beqn
\begin{array}{lcr}
    p_{1}\equiv j_{1} +k_{1} - i_{1}\;, &
 p_{2} \equiv j_{2} +k_{2} - i_{2}\;, &
q_{3} \equiv j_{2} +i_{2} - 1\;,  \\
   a_{12} \equiv p_{1} -1\;,  &   a_{23} \equiv  p_{2} -1 \;,
  &   a_{31} \equiv q_{3} -1\;,  \end{array} \;\;
\eeqn
 The inequalities on $i_{2}, j_{2}$  are found to be   equivalent to
  the following four inequalities:
\beqn
\label{eq:trieq}
 a_{12} +  a_{23} -  a_{31} &=& 2(k_{2}-i_{2}) \geq 0 \;\;\;. \nonumber \\
a_{12} - a_{23} +  a_{31} &=& 2(i_{2}-1) \geq 0 \;\;\;. \nonumber \\
  - a_{12} +  a_{23} +  a_{31} &=& 2(j_{2}-1) \geq 0 \;\;\;\nonumber \\
 a_{12} +  a_{23} +  a_{31} &=& 2(j_{2}+ k_{2}-2) \leq  2(m-2) \;\;.
\eeqn
 From  the  third and the fourth equation of eq.~(\ref{eq:trieq}), the
 inequality $ a_{12} \leq m-2$  follows, which is a condition for
${\cal F}_{2}( i_{1}=1, j_{1};k_{1}=1)$.
  Defining  a set
\beqn
\label{eq:D3}
{\cal D}_{3}(a_{1}, a_{2}, a_{3}) &\equiv&
   \{ (a_{1}, a_{2}, a_{3}) \mid \sum_{i(\neq j)}^{3} a_{i} - a_{j} \geq 0\;
\; {\rm for}\;\; i=1 \sim 3\; \;\;, \nonumber \\
 &~& \sum_{i=1}^{3} a_{i}
 = {\rm even} \leq 2(m-2) \} \;\;,
\eeqn
 we state  eq.~(\ref{eq:trieq})  as
\beqn
 {\cal F}_{3}(j_{1}, i_{2}, j_{2}) =
  {\cal D}_{3} (a_{12},a_{23},a_{31}) \;\;\;.
\eeqn
  We also write
\beqn
{\cal F}_{2}(j_{1}) \equiv {\cal F}_{2}( i_{1}=1, j_{1};k_{1}=1)
 \equiv {\cal D}_{2} (a_{12}) \;\;\;.
\eeqn
  for  the case $n=2$.

 Eq.~(\ref{eq:D3}) is nothing but the condition   that
 a triangle  be formed  which is made out of
 $a_{1},a_{2}$ and $a_{3}$  and whose  circumference is less than or equal
  to  $2(m-2)$.
  It is also the selection rule for the  three point function
  of the diagonal primaries in  $m$-th minimal unitary
 conformal field theory \cite{BPZ}.
In fact, the fusion rules for diagonal primary fields  read as
\beqn
 \langle \phi_{i i}~ \phi_{j j}~ \phi_{k k}\rangle
\not= 0 \quad,
\eeqn
if and only if
$i+j \geq k+1$ and two other permutations and $~i+j+k
{}~(=$ odd) $
\leq 2m-1$ hold.
This set of rules  is nothing but ${\cal D}_{3}(i-1, j-1, k-1)$.

  For the case $n=4$,  introduce
$p_{3} \equiv j_{3} +k_{3} - i_{3}\;,
  a_{34}\equiv p_{3} -1\;,
q_{4} \equiv j_{3} +i_{3} - 1\;,  a_{41}\equiv q_{4} -1 \;$.
   We find
\beqn
  {\cal F}_{2}( i_{3}, j_{3};k_{3}) =
  {\cal D}_{3} (a_{31},a_{34},a_{41}) \;\;\;\,
\eeqn
  The restrictions on the sum  in the case  $n=4$
  can be understood as gluing the two triangels:
\beqn
\label{eq:F4}
{\cal F}_{4}( j_{1}, i_{2}, j_{2}, i_{3}, j_{3}) &=&
 {\cal D}_{3} (a_{12},a_{23},a_{31})   \cap {\cal D}_{3} (a_{34},a_{41},a_{31})
  \nonumber \\
  &\equiv& {\cal D}_{4} (a_{12},a_{23},a_{34},a_{41}; a_{31})  \;\;.
\eeqn
 The allowed integers on $a_{31}$  are  naturally interpreted as  permissible
  quantum numbers flowing through an intermediate channel.
 As one can imagine, eq.~(\ref{eq:F4}) is not the only way to represent
 the restriction: one can also represent it as
\beqn
 &~&{\cal F}_{4}( j_{1}, i_{2}, j_{2}, i_{3}, j_{3})=
 {\cal D}_{3} (a_{12},a_{24},a_{41})
   \cap {\cal D}_{3} (a_{23},a_{34},a_{24}) \nonumber \\
 &\equiv& {\cal D}_{4} (a_{12},a_{23},a_{34},a_{41}; a_{24})  \;\;,
\eeqn
  which embodies the  crossing symmetric property of the amplitude.

 The restrictions in the  general case $n$ are understood as  attaching
  a  triangle   to  the case  $(n-1)$.
   To see this, define
\beqn
\begin{array}{lr}
  p_{\ell} = j_{\ell} + k_{\ell} - i_{\ell} \;,  &
  q_{\ell} = j_{\ell-1} +i_{\ell-1} -1 \;,  \\
 a_{\ell, 1} = q_{\ell}-1\;,  & a_{\ell, \ell+1} = p_{\ell}-1\;,
\end{array}  \;\;\; {\rm for}\;\;\;{\rm ~~\ell = 1,2, \cdots n} \;\;\;.
\eeqn
  Using $ 1 \leq i_{n-1} \leq k_{n-1},$~~
$1 \leq j_{n-1} \leq m - k_{n-1}$,   we derive
\beqn
 a_{n-1,n} +  a_{n,1} -  a_{n-1,1} &=& 2 (j_{n-1}-1) \geq 0 \;,
  \nonumber \\
a_{n-1,n} - a_{n,1} +  a_{n-1,1} &=& 2(k_{n-1} - i_{n-1}) \geq 0 \;,
 \nonumber \\
  - a_{n-1,n} +  a_{n,1} +  a_{n-1,1} &=& 2(i_{n-1}-1) \geq 0 \;,
 \nonumber \\
 a_{n-1,n} +  a_{n,1} +  a_{n-1,1}
&=& 2(j_{n-1}+ k_{n-1} -2) \leq  2(m-2) \;.
\eeqn
The restriction on $i_{n-1}$ and $j_{n-1}$  are, therefore,
  ${\cal D}_{3}(a_{n-1,n}, a_{n,1}, a_{n-1,1})$, which
 is what we wanted to see.
   All in all, we  find
\beqn
\label{eq:Dn}
 &~&{\cal F}_{n}( j_{1}, i_{2}, j_{2}, \cdots
  i_{n-1}, j_{n-1})   \nonumber \\
 &=&
  {\cal D}_{3}(a_{n-1,n}, a_{n,1}, a_{n-1,1}) \cap
 {\cal F}_{n-1}( j_{1}, i_{2}, j_{2}, \cdots
  i_{n-2}, j_{n-2})   \nonumber \\
 &=& {\cal D}_{3}(a_{n-1,n}, a_{n,1}, a_{n-1,1}) \cap
 {\cal D}_{n-1}(a_{1,2},a_{2,3}, \cdots a_{n-2,n-1}, a_{n-1,1};
  a_{3,1}, a_{4,1}, \cdots a_{n-2,1})  \nonumber \\
 &\equiv& {\cal D}_{n}(a_{1,2},a_{2,3}, \cdots a_{n-1,n}, a_{n,1};
  a_{3,1}, a_{4,1}, \cdots a_{n-1,1})
\eeqn

 From now on,  a shortened notation ${\cal D}_{n}(a_{1,2},a_{2,3},
 \cdots a_{n-1,n}, a_{n,1})$
 is understood  to represent
 $ {\cal D}_{n}(a_{1,2},a_{2,3}, \cdots a_{n-1,n}, a_{n,1};
  a_{3,1}, a_{4,1}, \cdots a_{n-1,1}) $.

  Putting eq~(\ref{eq:answer}) and eq.~(\ref{eq:Dn}) together, we obtain a
  formula
\beqn
\label{eq:fanswer}
 &~& P_{n}(\theta_{1}, \theta_{2}, \cdots \theta_{n})  \nonumber \\
 &=&
  - \left( \frac{ \partial \eta^{2}}{\partial \Lambda}\right)
 \left(\frac{-1}{\eta}\right)^{n}
 \sum_{ {\cal D}_{n}}  \left( \prod_{j = 2}^{n}
 \;\frac{\sinh (m- k_{j} -1)\theta_{j}
 }{\sinh m\theta_{j}  }  \right)
 \frac{\sinh (m-k_{1}-1)\theta_{1} }{\sinh m\theta_{1}}
 \;\;\;,
 \nonumber \\
\eeqn
 where ${\cal D}_n$ means  ${\cal D}_n(k_1-1,\cdots,k_n-1)$.
  Once again,  the  fact that
   the different divisions of ${\cal D}_{n}$ into $n-2$ triangles
   are   embodied  by   this  single expression  is precisely
  the statement of  the old duality.

  The object
  $P_{n}(\theta_{1}, \theta_{2}, \cdots \theta_{n})$
 is equipped with  $\theta_{j}$ and $k_j$  for $j=1,2,\cdots, n$
  and    any ${\cal D}_{3}(k_1-1,k_2-1,k_3-1)$ obeys the rule
 of  the triangle  specified above.
 It is, therefore,  natural to visualize    this as
  a vertex which  connects $n$ external legs corresponding  to $n$ loops.
  The vertex can be regarded as a dual graph of an n-gon that
 corresponds ${\cal D}_n(k_1-1,\cdots,k_n-1)$.

Using the formula   (\ref{eq:inverse}), we  perform the inverse Laplace
 transform  with respect to  $\zeta_{i}$ $(i=1, \sim n)$.
 We obtain
\beqn
\label{eq:fusionanswer}
&&\left( \prod_{j=1}^{n} {\cal L}_{j}^{-1} \right)
 \left( \frac{\Lambda}{N} \right)^{n-2} \left( \frac{\partial}
{\partial \Lambda} \right)^{n-3}
 \prod_{i=1}^{n}  \left( - \frac{\partial}{\partial (a \zeta_{i}) } \right)
 P_{n}(\theta_{1}, \theta_{2}, \cdots , \theta_{n})
  \nonumber \\
 &&\;\;
 = (-)^{n+1}
 \left( \frac{\Lambda}{N} \right)^{n-2} \left( \frac{\partial}
{\partial \Lambda} \right)^{n-3}
  \left[ \frac{\partial \eta^{2}}{\partial \Lambda}
  \left( \frac{1}{ a \eta }\right)^{n}
 \sum_{ {\cal D}_{n}} \prod_{j = 1}^{n}  \frac{ M \ell_{j}}{\pi}
  \underline{K}_{1-k_j/m} ( M\ell_{j}) \right] .
\eeqn
  where  $ {\cal L}_{j}^{-1}$  denotes  the inverse Laplace transform
  with respect to $\zeta_{j}$.
   Expressing this by $a$ and $t$ we obtain
\beqn
\label{eq:fusionanswerf}
&&  {\cal A}_{n}^{ fusion}( \ell_{1}, \ell_{2}, \cdots \ell_{n})
\nonumber \\
&&\;
 = - \frac{1}{m} \left( \frac{1}{m+1} \right)^{n-2}
  \left( \frac{\partial}{\partial t} \right)^{n-3}
 \left[ t^{-1 - \frac{(n-2)}{2m} } \sum_{ {\cal D}_{n}}  \prod_{j=1}^{n}
  \frac{M \ell_{j}}{\pi}  \underline{K}_{1-k_j/m} \left( M \ell_{j} \right)
 \right] .
\eeqn
  This is the answer quoted in the introduction.

   It is straightforward to  look at the small
 length behavior of
 eq.(\ref{eq:fusionanswerf}).
   This  was done in \cite{AII2}   in the case $n=3$,
 using
the formula
\beqn
\label{eq:Kform}
K_{\nu}(x) = \frac{\pi}{2 \sin(\nu \pi)}
\left[
\frac{1}{x^{\nu}} \left\{ \frac{2^{\nu}}{\Gamma(1-\nu)} + O(x^2)
                  \right\}
+ x^{\nu} \left\{ - \frac{1}{2^{\nu} \Gamma(1+\nu)} + O(x^2)
          \right\}
\right].
\eeqn
  The agreement with the approach from
 the generalized Kdv  flows \cite{DifK}
(See also \cite{AGBG}.)  has been given.
   We will not dwell on this point further.

\section{Residual Interactions }

Our formula  in the last section   tells how
  the higher order operators (gravitational descendants) in addition to the
dressed primaries  included in the form of the loop length are constrained to
  obey   the selection rules of CFT.
 The two-matrix model realizing  the unitary minimal series  coupled to
 gravity
  as the continuum limit of the $(m+1,m)$ symmetric critical point
  knows  the fusion rules and the  duality symmetry in the form of
  the loop operators.
  The term  we have dealt with in the last section for general $n$
  is, however, supplemented with an increasing number of
 other terms with $n$ $~(n\geq 4)$.  The existence of such terms itself
  implies that  the knowledge we obtain from  the two and the
 three point functions
   is not  sufficient to determine  the full amplitude for $n \geq4$.
  This coincides with  the notion of contact interactions familiar   from
 the field theory  of derivative couplings as
  well as in (super) string theory \cite{GrSe}.
    The counterpart of our  approach
  to this phenomenon  in the continuum framework  is presumably related to
   the discusson on the boundary of moduli space.


 In what follows, we show how to perform the inverse Laplace
 transformation of the resolvents to get loop
 amplitudes in terms of loop lengths in the case of $n=4$, $5$.
It is necessary to put
\beqn
  \tilde{\Delta}_n(\theta_1,\cdots,\theta_n)
  \equiv \tilde{\Delta}_n(z^*_1,\cdots,z^*_n) |_{\Lambda_i=\Lambda}
\eeqn
 in a manageable form to the inverse Laplace transform.
Let us recall that $P_n(\theta_1,\cdots,\theta_n)$ can be inverse
 Laplace transformed immediately.
 If $\tilde{\Delta}_n(\theta_1,\cdots,\theta_n)$ is expressed as
 a polynomial of $P_j(\theta_1,\cdots,\theta_n)$ and their derivatives
 with respect to $\Lambda$,
 the inverse Laplace transform  can be done immediately.
 Let us pursue this possibility.
We also make use of the fact that when one of the loops shrinks
 and the loop length goes to zero,
 the $n$-loop amplitude  must become proportional to the
 derivative of the $(n-1)$-loop amplitude with
 respect to   the cosmological constant.  We represent this fact by
\beq
 {\cal A}_n(\ell_1,\cdots,\ell_n) \to ~~\propto
 \frac{\partial}{\partial t} {\cal A}_{n-1}(\ell_1,\cdots,\ell_{n-1})
\quad.\eeq

The inverse Laplace transformation of $P_n(\theta_1,\cdots,\theta_n)$ is
\beqn
 {\cal L}^{-1} \left[ P_n(\theta_1,\cdots,\theta_n) \right]
 =-\Bigl( \frac{\partial\eta^2}{\partial\Lambda} \Bigr)
 \Bigl(\frac{-1}{\eta}\Bigr)^n \Bigl(\frac{M}{\pi}\Bigr)^n
 \;\sum_{ {\cal D}_n } \Bigl[ \prod_i \underline{K}_{1-k_i/m}(M\ell_i) \Bigl]
\eeqn
and in the limit  $M\ell\to 0$ we have
\beq
 \underline{K}_{1-k/m}(M\ell) \approx \frac{\pi 2^{-k/m}}{\Gamma(k/m)}
 (M\ell)^{k/m-1}
\quad.\eeq
 If  the n-th loop shrinks we have, therefore,
\beq
 {\cal L}^{-1}\left[ P_n(\theta_1,\cdots,\theta_n) \right]
 \to~~\propto
{\cal L}^{-1}\left[ P_{n-1}(\theta_1,\cdots,\theta_{n-1}) \right]
 \quad.
\label{eq:shrink-P}
\eeq
Then in the limit  $M\ell_n \to 0$,
 $\tilde{\Delta}_n(\theta_1,\cdots,\theta_n)$ must
 satisfy the following relation:
\beq
 {\cal L}^{-1}\left[ \tilde{\Delta}_{n}(\theta_1,\cdots,\theta_n) \right]
 \to~~\propto
 {\cal L}^{-1}\left[ \frac{\partial}{\partial \Lambda}
 \tilde{\Delta}_{n-1} (\theta_1,\cdots,\theta_{n-1})  \right]
\label{eq:shrink-Delta}
\eeq
This relation restricts the possible form of
 $\tilde{\Delta}_n(\theta_1,\cdots,\theta_n)$ .

As we want $\tilde{\Delta}_n(\theta_1,\cdots,\theta_n)$ to be expressed as
 a polynomial of $P_j$ and their derivatives with respect to $\Lambda$,
 we need to introduce  a notation
\beqn
\label{eq:notationS}
&~& \left[ {\cal S} P_{1, \cdots, i_{1}} P_{j_{2}, \cdots, j_{2}+ i_{2}-1}
  \cdots P_{n-i_{\ell}+1, \cdots, n} \right] \left( \theta_{1}, \theta_{2},
 \cdots \theta_{n} \right) \equiv   \\
 &~& \frac{1}{n!} \sum_{\sigma \in  {\cal P}_{n}}
  P_{i_{1}} (\theta_{\sigma(1)}, \cdots,
 \theta_{\sigma(i_{1})}) P_{i_{2}}(\theta_{\sigma(j_{2})},
 \cdots, \theta_{\sigma(j_{2} +i_{2}-1)} )
  \cdots  P_{i_{\ell}}(\theta_{\sigma(n-i_{\ell} +1)},
 \cdots, \theta_{\sigma(n)} ) \;, \nonumber
\eeqn
where ${\cal P}_n$ represents   the permutations of $(1,2,\cdots,n)$.
 To be more specific,  for example
\beqn
\label{eq:SP}
\left[{\cal S}P_{123}P_{234} \right]  (\theta_{1}, \theta_{2},
\theta_{3},\theta_{4})
  = \frac{1}{4!} \sum_{\sigma \in {\cal P}_{4}}
 P_{3}(\theta_{\sigma(1)}, \theta_{\sigma(2)},\theta_{\sigma(3)})
 P_{3}(\theta_{\sigma(2)}, \theta_{\sigma(3)},\theta_{\sigma(4)})
 \nonumber \\
\left[{\cal S}P_{1234}P_{34} \right] (\theta_{1}, \theta_{2},
\theta_{3},\theta_{4})
  = \frac{1}{4!} \sum_{\sigma \in {\cal P}_{4}}
 P_{4}(\theta_{\sigma(1)}, \theta_{\sigma(2)},\theta_{\sigma(3)}
 \theta_{\sigma(4)} )
 P_{2}(\theta_{\sigma(3)},\theta_{\sigma(4)}) \;\;\;.
\eeqn

 It is convenient to  represent $P_n(\theta_1,\cdots,\theta_n)$ by
 an n-vertex which connects n external legs.
For example for $n=2$, 3 and 4
\beqn
  &&P_2(\theta_1,\theta_2) \equiv
 \epscenterboxy{2.5cm}{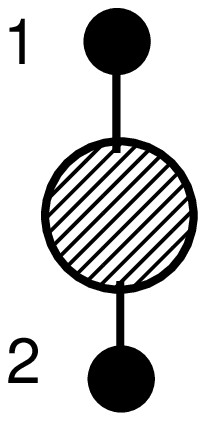}  \quad,\quad
  P_3(\theta_1,\theta_2,\theta_3) \equiv \epscenterboxy{2.5cm}{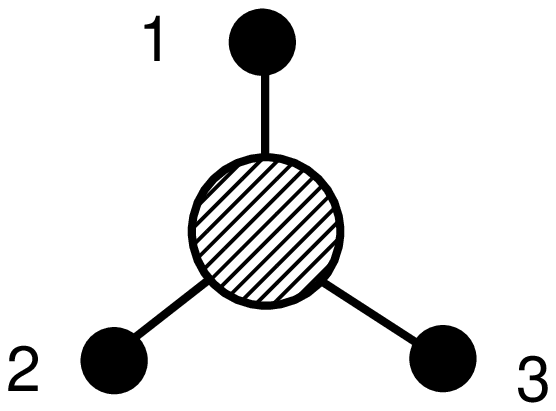}
  \nonumber \\
  &&P_4(\theta_1,\theta_2,\theta_3,\theta_4) \equiv
 \epscenterboxy{2.5cm}{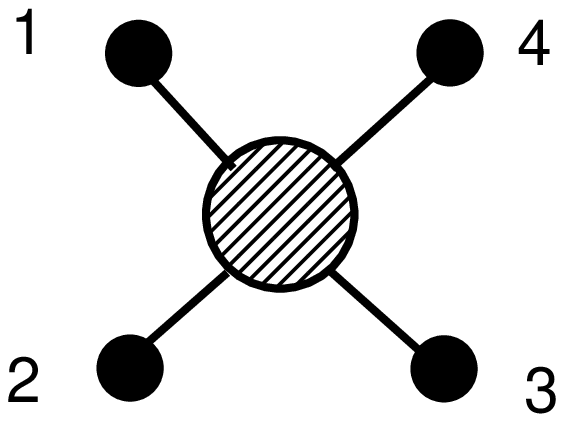}
\eeqn
The n-vertex  can be regarded as a dual graph of the n-gon (polygon)
 which corresponds to ${\cal D}_n$.
In terms of these vertices, let us express eq.~(\ref{eq:SP}) as follows.
\beqn
\left[{\cal S}P_{123}P_{234} \right]
 (\theta_{1}, \theta_{2},\theta_{3},\theta_{4})
  &\equiv& \epscenterboxy{2.5cm}{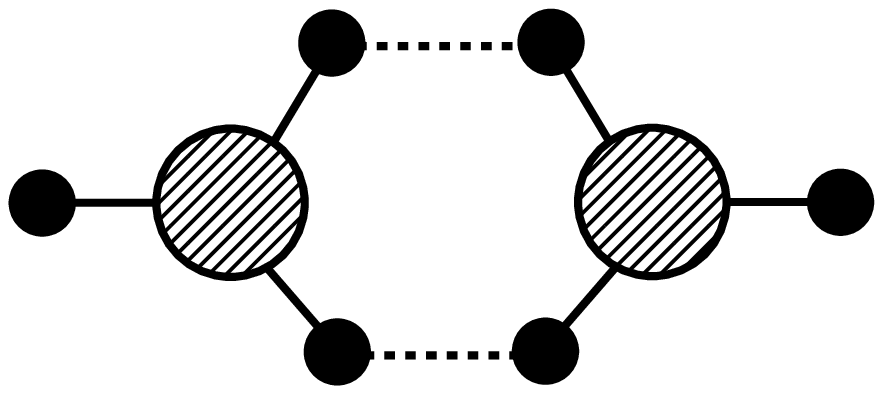}
  \nonumber \\
\left[{\cal S}P_{1234}P_{34} \right] (\theta_{1}, \theta_{2},\theta_{3},
\theta_{4})  &\equiv& \epscenterboxy{2.5cm}{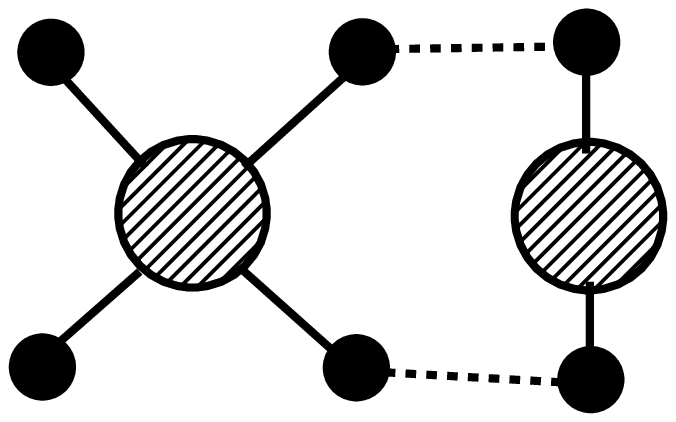}
\eeqn

The relation eq.~(\ref{eq:shrink-P}) can be represented, for example,  as
\beq
  \epscenterboxy{2.5cm}{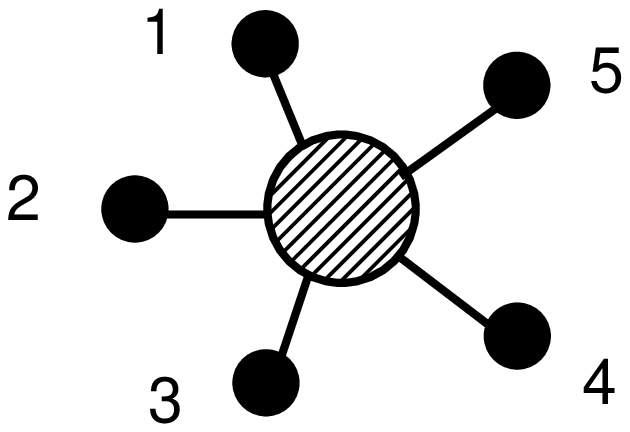}\qquad\to \quad\propto \qquad
  \epscenterboxy{2.5cm}{ai_fig/P1234.eps}
\eeq
 in the case of n=5.

Now we are concerned with the case of n=4 first.
Let us recall that for n=3
\beq
 \tilde{\Delta}_3(\theta_1,\theta_2,\theta_3)
 \propto P_3(\theta_1,\theta_2,\theta_3)
\quad.\eeq
$\tilde{\Delta}_4(\theta_{1}, \theta_{2}, \theta_{3}, \theta_{4})$ must
 include a
 term which becomes proportional to
 $P_3(\theta_{1}, \theta_{2}, \theta_{3})$
 in the limit $M\ell_4\to 0$,  which is
 $P_4(\theta_{1}, \theta_{2}, \theta_{3}, \theta_{4})$.
$\tilde{\Delta}_4(\theta_{1}, \theta_{2},\theta_{3}, \theta_{4})$ may
 also include terms which vanish in this limit. Such terms must consist
 of the product of two  multi-vertices which have 6 external legs in total.

  By explicit computation, we find
\beqn
  && \frac{-1}{2!} \tilde{\Delta}_{4}(\theta_{1}, \theta_{2},
  \theta_{3}, \theta_{4}) \mid  =
  \frac{\partial}{\partial \Lambda} P_{4}
(\theta_{1}, \theta_{2},\theta_{3}, \theta_{4})
 -\left[ {\cal S} P_{123}P_{234} \right] (\theta_{1}, \theta_{2},\theta_{3},
 \theta_{4})
   \nonumber \\
 &&\qquad +
 \left[ {\cal S} P_{1234}P_{34} \right]  \left(\theta_{1}, \theta_{2},
\theta_{3}, \theta_{4} \right)
\nonumber \\
&&\qquad =\epscenterboxy{2.5cm}{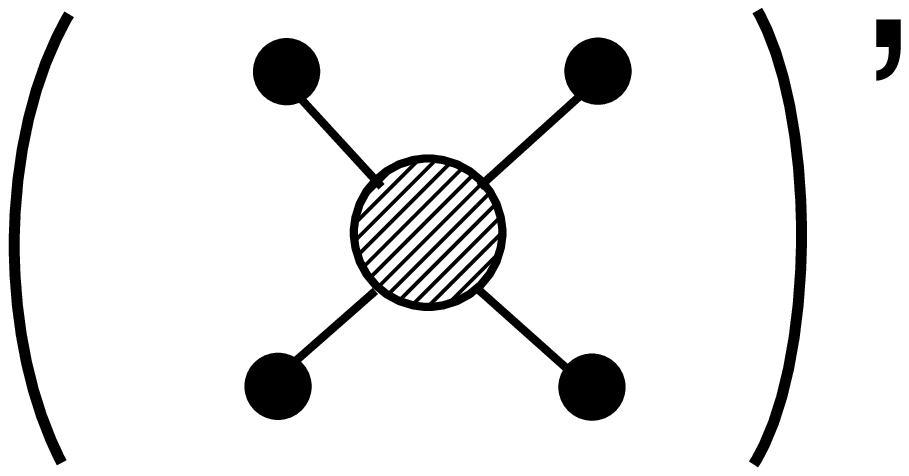}
\nonumber \\
&&\qquad + \epscenterboxy{2.5cm}{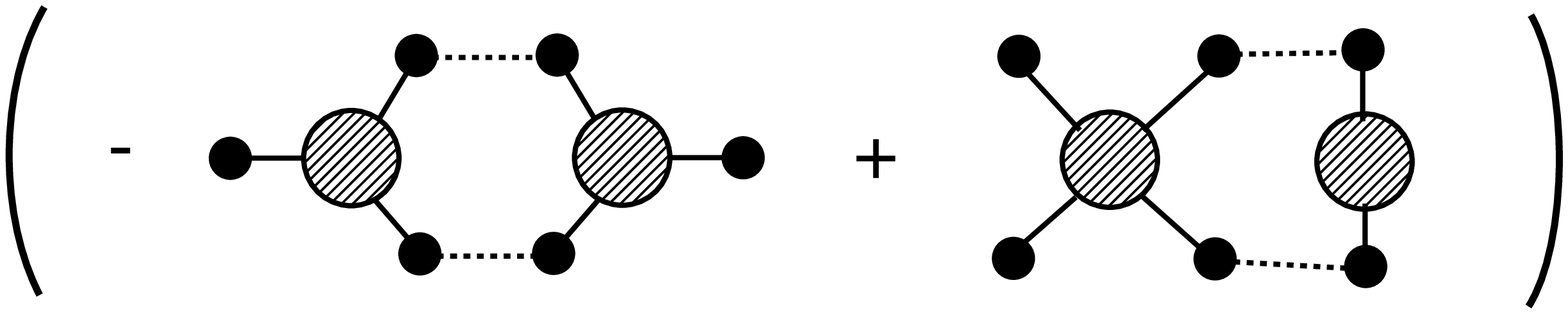}
\quad ,
\label{eq:n=4Delta}
\eeqn
where the prime represents the differentiation with respect to $\Lambda$.

  Carrying out the procedure indicated in (\ref{eq:procedure})  for the case
 $n=4$ together with eqs.~(\ref{eq:result}),  (\ref{eq:n=4Delta}), we obtain
  the  complete answer for  the macroscopic  four loop ampitude:
\footnote{  This was briefly reported in \cite{talk}.}
\beqn
\label{eq:n=4answer}
&~& {\cal A}_{4}( \ell_{1}, \cdots \ell_{4}) = {\cal A}_{4}^{fusion}
( \ell_{1} \cdots \ell_{4})  \nonumber \\
&+&  \frac{1}{m^{2}(m+1)^{2} } t^{-2 -\frac{1}{m} }
\left( \prod_{j=1}^{4} \frac{ M \ell_{j} }{\pi} \right)
 \left( \frac{1}{4!} \sum_{\sigma \in {\cal P}_{4}} \right)
\left\{ - \sum_{ {\cal D}_{3}}  \sum_{ {\cal D}_{3}^{\prime} }
\prod_{j=1}^{4} \left[ B_{j}^{123} \ast  B_{j}^{\prime 234}\right]
\left( M \ell_{\sigma(j)} \right)  \right.
  \nonumber \\
&+&
\left. \sum_{ {\cal D}_{4}}  \sum_{ {\cal D}_{2}^{\prime} }
\prod_{j=1}^{4} \left[ B_{j}^{1234} \ast  B_{j}^{\prime 34}\right]
\left( M \ell_{\sigma(j)} \right)
\right\}  \;\;\;.
\eeqn
  Here $B_{j}^{123} = (  \underline{K}_{1-k_1/m},
 \underline{K}_{1-k_2/m},  \underline{K}_{1-k_3/m},
1  )$, $B_{j}^{\prime 234} =
 ( 1, \underline{K}_{1-k'_2/m},
 \underline{K}_{1-k'_3/m},
  \underline{K}_{1-k'_4/m})$,
$B_{j}^{1234}$ $ = (  \underline{K}_{1-k_1/m},
 \underline{K}_{1-k_2/m},  \underline{K}_{1-k_3/m},
 \underline{K}_{1-k_4/m}  )$ and
 $B_{j}^{\prime 34} =
 ( 1, 1, \underline{K}_{1-k'_3/m},
  \underline{K}_{1-k'_4/m})$.
   We have  introduced
 ${\cal D}_{3} \equiv {\cal D}_{3}(k_1-1,k_2-1,k_3-1)$,
 $ {\cal D}_{3}^{\prime}
 \equiv {\cal D}_{3}(k'_1-1,k'_2-1,k'_3-1) $,
  $ {\cal D}_{4}
\equiv  {\cal D}_{4}(k_1-1,k_2-1,k_3-1,k_4-1)$
 and
  $ {\cal D}_{2}^{\prime}  \equiv  {\cal D}_{2}
(k'_1-1,k'_2-1 )$  and
 have defined  the convolution  $A\ast B(M\ell)$ by
\beqn
   \left[ A\ast B \right]  (M\ell)  \equiv  \int_{0}^{\ell}
 \frac{ M  d \ell^{\prime} }{\pi}
 A(M\ell^{\prime}) B(M (\ell -\ell^{\prime}) ) \;\;\;.
\eeqn

Let us now turn to the n=5 case.
$\tilde{\Delta}_5(\theta_{1},\cdots,\theta_{5})$ include a term
 which is proportional to
\beq
\label{eq:n=5type1}
  \frac{\partial^2} {\partial \Lambda^2}P_5(\theta_{1},\cdots,\theta_{5})
  =\epscenterboxy{2.5cm}{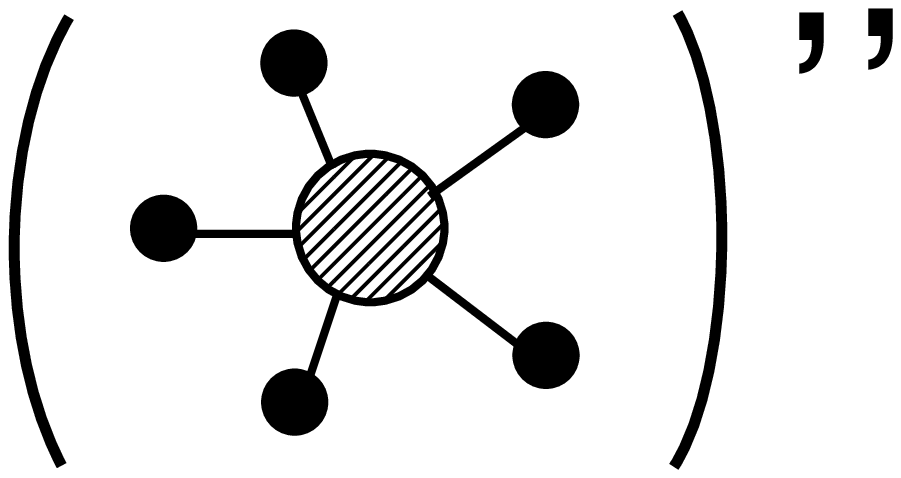}
\eeq
corresponding to the first term in eq.~(\ref{eq:n=4Delta}).
Corresponding to the second term in eq.~(\ref{eq:n=4Delta}),
 $\tilde{\Delta}_5(\theta_{1},\cdots,\theta_{5})$ must include a term
 which  consists of the product of two multi-vertices with 7 extenal
 legs in total.
The possible form is
\beqn
\label{eq:n=5type2}
 &&a \Bigl[{\cal S}\frac{\partial}{\partial \Lambda}(P_{12345})P_{45} \Bigr]
 +b \Bigl[{\cal S}P_{12345}\frac{\partial}{\partial \Lambda}(P_{45}) \Bigr]
 \nonumber \\
 &&\qquad+c \Bigl[{\cal S}\frac{\partial}{\partial \Lambda}
(P_{1234})P_{345} \Bigr]
 +d \Bigl[{\cal S}P_{1234}\frac{\partial}{\partial \Lambda}(P_{345}) \Bigr]
 \nonumber \\
 && = a \epscenterboxy{2.5cm}{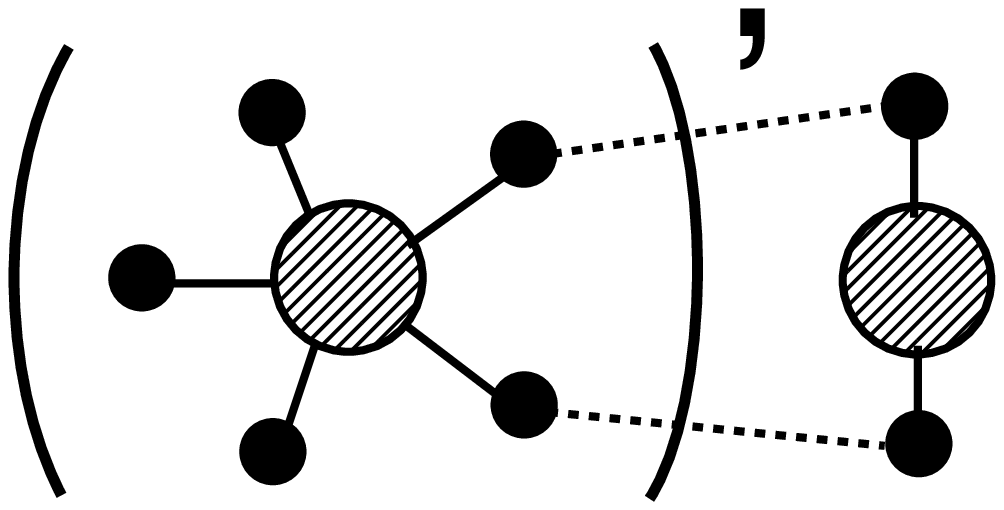}
 +b \epscenterboxy{2.5cm}{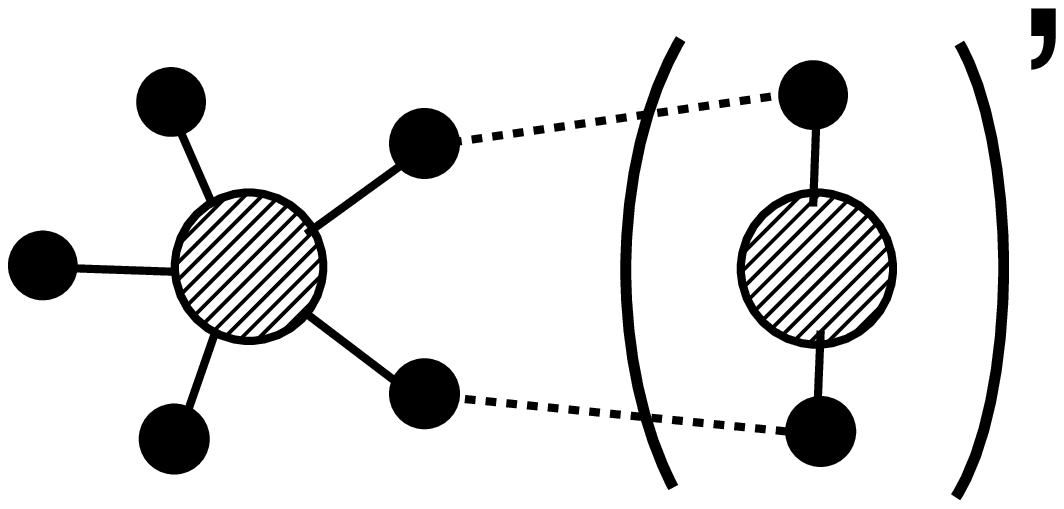}
 \nonumber \\
 &&\qquad + c \epscenterboxy{2.5cm}{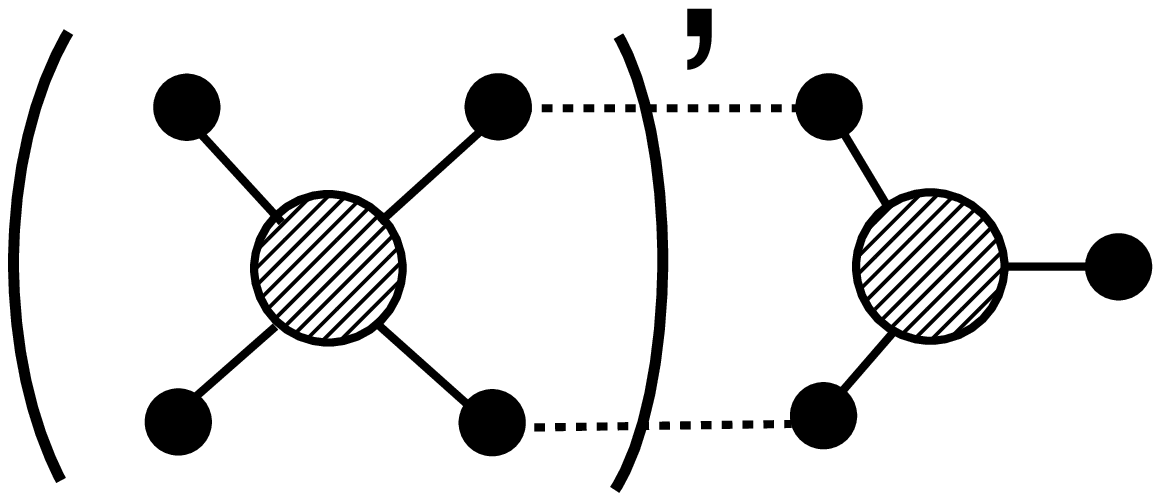} +
 d \epscenterboxy{2.5cm}{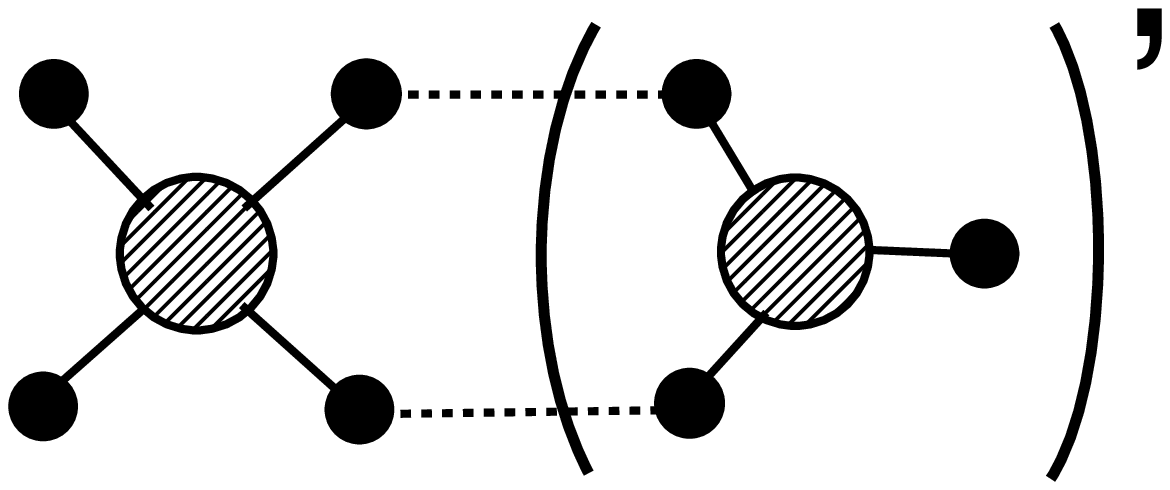}
\quad.
\eeqn
In the limit  $M\ell_5\to 0$, it becomes
\beqn
 (3a+c) \Bigl[{\cal S}\frac{\partial}{\partial \Lambda}(P_{1234})P_{34} \Bigr]
 +(3b+d) \Bigl[{\cal S}P_{1234}\frac{\partial}{\partial \Lambda}(P_{34}) \Bigr]
 \nonumber \\
 +2c \Bigl[{\cal S}\frac{\partial}{\partial \Lambda}(P_{123})P_{234} \Bigr]
 +2d \Bigl[{\cal S}P_{123}\frac{\partial}{\partial \Lambda}(P_{234}) \Bigr]
\quad.\eeqn
  We  require  this expression to be proportional to the $\Lambda$-derivative
 of  the second term
 in eq~.(\ref{eq:n=4Delta}).  We find
\beq
a=(1-c)/3,\quad b=(2+c)/3,\quad d=-1-c
\quad .
\eeq
$\tilde{\Delta}_5(\theta_{1},\cdots,\theta_{5})$ may include terms
 which vanish in the limit under consideration as well.
 They must  consist of the products of three multi-vertices with 9 external
 legs in total.
As one of the such terms we have
\beqn
\label{eq:n=5type3}
&& [{\cal S}P_{1234}P_{34}P_{45}] -[{\cal S}P_{1234}P_{23}P_{345}]
 +[{\cal S}P_{123}P_{124}P_{235}]
\nonumber \\
&&=\epscenterboxy{2.5cm}{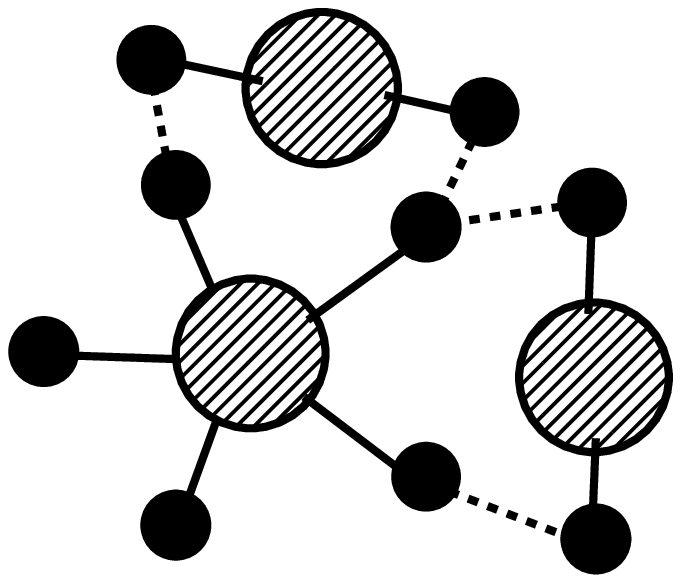}
-2 \epscenterboxy{2.5cm}{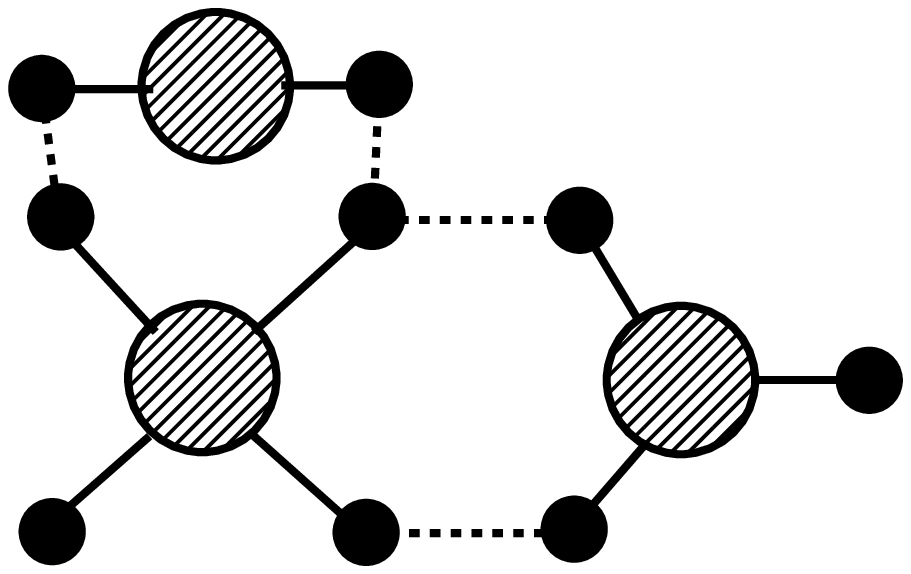}
\nonumber \\
&&\qquad +\epscenterboxy{2.5cm}{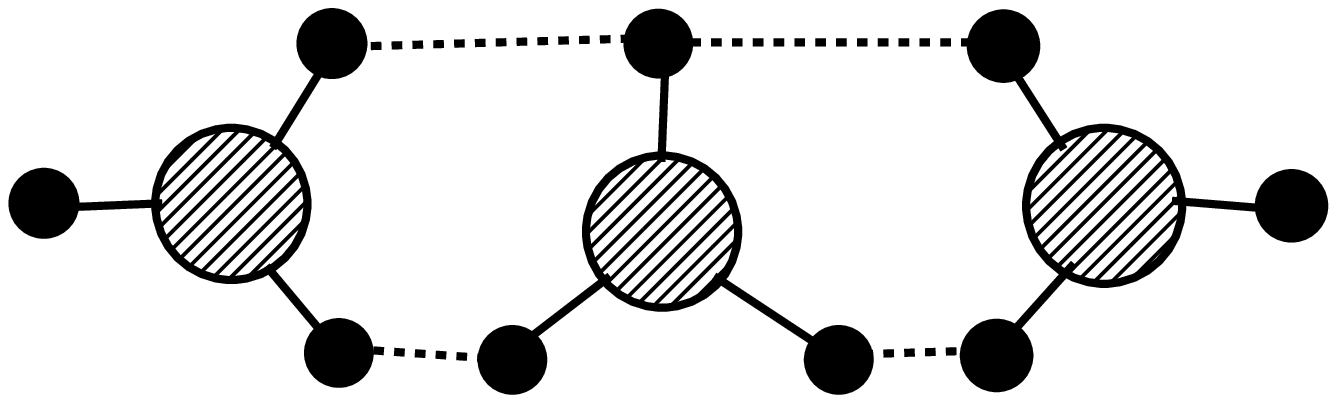}
\quad .
\eeqn
( There are some other combinations which satisfy the conditions. )
$\tilde{\Delta}_5(\theta_{1},\cdots,\theta_{5})$ must be expressed as a
 linear combination of the above three types of terms eq.~(\ref{eq:n=5type1}),
  eq.~(\ref{eq:n=5type2}) and eq.~(\ref{eq:n=5type3}) if
  the assumption under consideration is true.
By explicit calculation, we have found in fact that
 $\tilde{\Delta}_5(\theta_{1},\cdots,\theta_{5})$ can be expressed
 as a linear combination of
 eq.~(\ref{eq:n=5type1}),  eq.~(\ref{eq:n=5type2})
 and eq.~(\ref{eq:n=5type3}) :
\beqn
 &&\frac{-1}{3!}\tilde{\Delta}_5(\theta_{1}, \theta_{2},  \theta_{3},
 \theta_{4}, \theta_{5}) =
 \left( \frac{\partial}{\partial \Lambda} \right)^{2}
 P_{5}(\theta_{1}, \theta_{2},  \theta_{3}, \theta_{4}, \theta_{5})
  \nonumber \\
 &+&\left[{\cal S}P_{12345} \left(
 2\frac{\stackrel{\rightarrow}{\partial} }
{\partial \Lambda}
 + 3\frac{\stackrel{\leftarrow}{\partial} }
{\partial \Lambda}
 \right) P_{45}  \right]
(\theta_{1}, \theta_{2},  \theta_{3}, \theta_{4}, \theta_{5})
 \nonumber \\
 &-& \left[{\cal S}  P_{1234} \left(
 \frac{ \stackrel{\rightarrow}{\partial} }{\partial \Lambda} +4
 \frac{ \stackrel{\leftarrow}{\partial} }{\partial \Lambda}
 \right) P_{345}  \right]
  (\theta_{1}, \theta_{2},  \theta_{3}, \theta_{4}, \theta_{5})  \nonumber \\
 &+& \left[ \left( 2P_{1234}P_{34}P_{45} -4P_{123}P_{23}P_{345}
 +2P_{123}P_{124}P_{235} \right) \right]
 (\theta_{1}, \theta_{2},  \theta_{3}, \theta_{4}, \theta_{5})\;.
\eeqn

  Following the same procedure as obtaining eq.~(\ref{eq:n=4answer}),
  we find  the complete answer for
  the five loop amplitude:
\beqn
\label{eq:n=5answer}
&~& {\cal A}_{5}( \ell_{1}, \cdots \ell_{5}) = {\cal A}_{5}^{fusion}
( \ell_{1} \cdots \ell_{5}) +
 \frac{1}{m^{2}(m+1)^{3} }
\left( \prod_{j=1}^{5} \frac{ M \ell_{j} }{\pi} \right)
 \left( \frac{1}{5!} \sum_{\sigma \in {\cal P}_{5}} \right)  \nonumber \\
&\times& \left[
 \left( 2  t^{-1 -\frac{3}{2m}} \left( \frac{\partial}{\partial t}
  t^{-1} \right)_{R}
  + 3 t^{-1} \left( \frac{\partial}{\partial t}   t^{-1 -\frac{3}{2m}}
 \right)_{L} \right)
  \sum_{ {\cal D}_{5}}  \sum_{ {\cal D}_{2}^{\prime} }
 \prod_{j=1}^{5} \left[
  B_{j}^{12345} \ast  B_{j}^{\prime 45} \right]
(M \ell_{\sigma(j)} )  \right.   \nonumber \\
&-&  \left(  t^{-1-\frac{1}{m}}
 \left( \frac{\partial}{\partial t}   t^{-1-\frac{1}{2m}} \right)_{R}
  + 4  t^{-1-\frac{1}{2m}} \left( \frac{\partial}{\partial t}
  t^{-1-\frac{1}{m}} \right)_{L}
 \right)
  \sum_{ {\cal D}_{4}}  \sum_{ {\cal D}_{3}^{\prime} }
 \prod_{j=1}^{5} \left[ B_{j}^{1234} \ast
  B_{j}^{\prime 345} \right]
(M \ell_{\sigma(j)} )      \nonumber \\
&-&
  \frac{1}{m } t^{-3- \frac{3}{2m}}
\left( 2\sum_{ {\cal D}_{5}}  \sum_{ {\cal D}_{2}^{\prime} }
 \sum_{ {\cal D}_{2}^{\prime \prime} }
 \prod_{j=1}^{5} \left[ B_{j}^{12345}
    \ast B_{j}^{\prime 34} \ast B_{j}^{\prime \prime 45} \right]
(M \ell_{\sigma(j)} )   \right.   \nonumber \\
&-& 4\sum_{ {\cal D}_{4}}  \sum_{ {\cal D}_{2}^{\prime} }
 \sum_{ {\cal D}_{3}^{\prime \prime} }
 \prod_{j=1}^{5} \left[ B_{j}^{1234}
    \ast B_{j}^{\prime 23} \ast B_{j}^{\prime \prime 345} \right]
(M \ell_{\sigma(j)} )     \nonumber \\
  &+&  \left.  2\sum_{ {\cal D}_{3}}  \sum_{ {\cal D}_{3}^{\prime} }
 \sum_{ {\cal D}_{3}^{\prime \prime} }
   \left. \prod_{j=1}^{5} \left[ B_{j}^{123}
    \ast B_{j}^{\prime 124} \ast B_{j}^{\prime \prime 235} \right]
(M \ell_{\sigma(j)} )   \right)  \right]     \;\;\;.
\eeqn
  Here
 $\left( \frac{\partial}{\partial t} \right)_{L,(R)} $ means
  the derivative acting only on the left(right) part of the
 convolutions.  The rest of the notations here are similar to
   those of the $n=4$  case and will be self-explanatory.
\footnote{
  It is   just  a matter of writing  to take a small length limit
  of eqs.~(\ref{eq:n=4answer}),(\ref{eq:n=5answer}) to obtain
  the corresponding expression for the microscopic operators. (Use
 eq.~(\ref{eq:Kform})). }

We conjecture that  $\tilde{\Delta}_n(\theta_{1},\cdots,\theta_{n})$
 can be represented as a polynomial of $P_j(\theta_{1},\cdots,\theta_{j})$ and
 their $t$ derivatives for any $n$  :  the  final answer
 would  then be obtained   by  convolutions  of various $B_{j}$'s
 and their derivatives.
If the conjecture is true in fact,  the power counting argument
  tells us that  the j-th term of $\tilde{\Delta}_n$ turns out to be
  represented by a figure which
 consists of the products of j  multi-vertices with $n+2j$ external legs
 in total.
We  hope that, for higher loops, $\tilde{\Delta}_n(\theta_{1},\cdots,
\theta_{n})$ can be
 put   in principle in  a form  such as eqs.~(\ref{eq:n=4answer}),
(\ref{eq:n=5answer}) in the same manner
  as we have determined  the four and five loops from the lower ones.
Explicit determination  of the full amplitude in this way beyond
   five loops, however, appears to us  still very formidable.


\section{Acknowledgements}
  We thank Atsushi Ishikawa for enjoyable collaboration in
 \cite{AII,AII2}.
   We also thank Kenji Hamada, Keiji Kikkawa, and Alyosha Morozov
  for helpful discussions on this subject.


\newpage
\appendix
\section{A}
  In this  appendix, we  prove the recursion relations for
\beqn
  \left( \begin{array}{clcr}
 m_{1}, & m_{2}, & m_{3}, & \cdots \\
 i_{1}, & i_{2}, & i_{3}, &  \cdots
 \end{array} \right)_{n}   \;\;\;\;
  \;\; {\rm with}\;\;\;\; \sum_{\ell} m_{\ell}  \leq  n-2 \;\;\;
\eeqn
 introduced in the text.  The proof goes by  mathematical
 inductions.

   We will first prove  the simplest case
\beqn
\label{eq:simpler1}
 \left( \begin{array}{cccc}
 m, & 0, & \cdots, & 0 \\
 i_{1},  & i_{2}   & \cdots, & i_{n}
 \end{array} \right)_{n}   =
 \frac{1}{ {\displaystyle \prod_{j (\neq i_{1})}^{n} [i_1-j]^{2} } }
 \quad,\qquad {\rm for}\quad m=n-2
\eeqn
and
\beqn
\label{eq:simpler2}
 \left( \begin{array}{cccc}
 m, & 0, & \cdots, & 0 \\
 i_{1},  & i_{2}   & \cdots, & i_{n}
 \end{array} \right)_{n}
 =   0
 \quad,\qquad {\rm for}\quad m\le n-3 \quad.
\eeqn

  Assume that eq.~(\ref{eq:simpler1})  and eq.~(\ref{eq:simpler2}) are true
 at $n$.
  Without loss of generality, $i_{1}$ can be taken to be $1$.
Let us consider the left hand side of eq.~(\ref{eq:simpler1}) or
 eq.~(\ref{eq:simpler2})
 in which $n$ is replaced by $n+1$.
To compute  them we  observe that   the elements of ${\cal S}_{n+1}$  are
 generated
  by associating
  $n$  different ways of
 inserting  $[n+1]$    with each element $\sigma \in {\cal S}_{n}$.
  In the case where $[n+1]$ is inserted in between $[1]$ and $[\sigma(1)]$,
   this contribution  is equal to
\beqn
\label{eq:cont1}
  - \sum_{\sigma \in {\cal S}_{n} }
   \frac1{[1-\sigma(1)]^m} \; \frac{[1- \sigma(1)]^{m}}{ [1- (n+1)]^{m+1}
 [(n+1)- \sigma(1)] } \prod_{j=2}^{n}
 \frac1{[j - \sigma(j) ]} \;\;\;.
\eeqn
  The contributions   from  the sum of the remaining $n-1$  insertions
  are  found to be  equal to
\beqn
\label{eq:cont2}
  -  \sum_{\sigma \in {\cal S}_{n} }
  \frac1{[1-\sigma(1)]^m} \; \frac{1}{ [\sigma(1)- (n+1)]
 [1 -(n+1)] } \prod_{j=2}^{n} \frac1{[j - \sigma(j) ]} \;\;\;.
\eeqn
  Here we have used
\beqn
\label{eq:frac}
\frac{1}{[j-(n+1)] [(n+1)-m]}  =  \frac{1}{[j-m]} \left(
  \frac{1}{[j-(n+1)]} -
 \frac{1}{[m-(n+1)]} \right)\;\;\;.
\eeqn

 Note also that $ {\displaystyle\prod_{j=1}^{n-1} }
 1/[\sigma^{j}(1) -\sigma^{j+1}(1) ]
 = {\displaystyle \prod_{j=2}^{n} } 1/[j - \sigma(j) ]$.
  Putting eqs.~(\ref{eq:cont1})  and (\ref{eq:cont2}) together,
  we find
\beqn
\label{eq:**}
  &~&  \left( \begin{array}{cccc}
 m, & 0, & \cdots, & 0 \\
 1, & 2, & \cdots, & n+1
 \end{array} \right)_{n+1}
  =  \nonumber \\
   ~&-&  \sum_{\sigma \in {\cal S}_{n} }
  \frac1{[1-\sigma(1)]^m} \;
 \frac{  \left \{ 1-  \left( \frac{[1- \sigma(1)]}{[1-(n+1)]} \right)^{m}
 \right \}  } { [\sigma(1)- (n+1)] [1 -(n+1)] }
 \prod_{j=2}^{n} \frac1{[j - \sigma (j) ]} \;\;\;.
\eeqn
  Factorizing the expression inside the bracket
 $\left \{ ~~ \cdots ~~  \right \}$, we have
\beq
\label{eq:recursion1}
 \left( \begin{array}{cccc}
 m, & 0, & \cdots, & 0 \\
 1, & 2, & \cdots, & n+1
 \end{array} \right)_{n+1}
  =  \nonumber \\
 \sum_{l=1}^{m-1}\left( \begin{array}{cccc}
 m-l, & 0, & \cdots, & 0 \\
 1,   & 2, & \cdots, & n
 \end{array} \right)_{n}  \frac{1}{[1-(n+1)]^{1+l} } \quad.
\eeq
Then from the assumption, eq.~(\ref{eq:simpler1}) and eq.~(\ref{eq:simpler2})
 are also satisfied when $n$ is replaced by $n+1$.
On the other hand for $n=3$ eq.~(\ref{eq:simpler1}) and
 eq.~(\ref{eq:simpler2}) are
 clearly true, so we have proven the relations.

  Now we turn to the more general case the proof of which is a straightforward
 generalization of the one given above.
  To derive the relations
\beqn
 \label{eq:relation3}
 \left( \begin{array}{cccccc}
 m_{1}, & \cdots, & m_{k}, & 0,       & \cdots, & 0 \\
 i_{1}, & \cdots  & i_{k}, & i_{k+1}, & \cdots, & i_{n}
 \end{array} \right)_{n}
 =0 \quad,\qquad {\rm for} \quad
    \sum_{\ell=1}^{k} m_{\ell}\le n-3 \quad,
\eeqn
and
\beqn
 &&\label{eq:relation4}
 \left( \begin{array}{cccccc}
 m_{1}, & \cdots, & m_{k}, & 0,          & \cdots, & 0 \\
 i_{1}, & \cdots  & i_{k}, & \ i_{k+1},  & \cdots, & i_{n}
 \end{array} \right)_{n} \nonumber \\
 &&\qquad =\sum_{j=1}^{k}
 \left( \begin{array}{cccccccc}
 m_{1}, & \cdots, & m_{j}-1, & \cdots, &m_{k}, & 0, & \cdots, & 0 \\
 i_{1}, & \cdots, & i_{j},   &\cdots,  &i_{k}, & i_{k+1}, & \cdots, & i_{n-1}
 \end{array} \right)_{n-1} \frac{1}{[j-n]^2}
 \quad, \nonumber \\
 &&\qquad\qquad\qquad\qquad\qquad\qquad\qquad\qquad\qquad
 {\rm for}\quad \sum_{\ell=1}^{k} m_{\ell}=n-2
 \quad.
\eeqn
Let us assume eq.~(\ref{eq:relation3}) at $n$ .

 We take  $ i_{\ell} = \ell,~~$ $ \ell =1 \sim k $
  without  loss of generality.  The way in which   the elements of
 ${\cal S}_{n+1}$ are generated is the same as the one given above.
  In the case where $[n+1]$ is inserted in between $[\ell]$ and
 $[\sigma \left( \ell \right)]$ $ \ell = 1 \sim k$,
  the contribution is
\beqn
\label{eq:cont1g}
 -  \sum_{\sigma \in {\cal S}_{n} }
  \frac1{[1-\sigma(1)]^{m_1+1}} \cdots
  \frac1{[\ell-\sigma(\ell)]^{m_{\ell}}} \;
  \frac{[\ell- \sigma(\ell)]^{m_{\ell} } }
 { [\ell- (n+1)]^{m_{\ell}+1} [(n+1)- \sigma(\ell)] } \;\;\;  \nonumber \\
  \times  \frac1{ [ (\ell+1) -\sigma(\ell +1)]^{m_{\ell+1}+1}}
 \cdots  \frac1{ [k-\sigma(k)]^{m_{k}+1}}
 \prod_{j ( \neq 1 , 2, \cdots k)}^{n-1}
 \frac1{[j - \sigma (j) ]} \;\;\;.
\eeqn
  The contributions   from  the sum of the remaining   insertions
  are
\beqn
  &-&  \sum_{\sigma \in {\cal S}_{n} }
  \sum_{\ell ( \neq  p_{2},\cdots p_{k} )}^{n-1}
  \frac1{ [1-\sigma(1)]^{m_{1}+1 } } \cdots  \frac1{ [k-\sigma(k)]^{m_{k+1} }}
  \nonumber \\
 &\times&
   \prod_{j (\neq p_{2}, \cdots p_{k}) }^{n-1}
   \frac1{[\sigma^{j}(1) - \sigma^{j+1}(1) ]}
   \left(  \frac{1}{[ \sigma^{\ell}(1) -(n+1)]}
  - \frac{1}{[ \sigma^{\ell +1}(1) -(n+1)]} \right) \;\;\;.
\eeqn
  Here  $p_{\ell}$  $\ell = 1 \sim k$ are such that
  $ \sigma^{p_{\ell}} (1) = \ell$.
  Using  eq. ~(\ref{eq:frac}) again,  we find that this equals
\beqn
\label{eq:cont2g}
  -  \sum_{\sigma \in {\cal S}_{n} } \sum_{\ell =1}^{k}
  \frac1{[ 1-\sigma(1)]^{m_{1}+1}}  \cdots
  \frac1{ [ \ell-\sigma(\ell)]^{m_{\ell}}}\;  \frac{1}{ [\sigma(\ell)- (n+1)]
  [\ell -(n+1)] }  \nonumber \\
  \frac1{[ (\ell +1) -\sigma(\ell + 1)]^{m_{\ell+1}+1}}
  \cdots  \frac1{[ k -\sigma(k)]^{m_{k}+1}}
  \prod_{j (\neq 1, 2, \cdots k)}^{n-1}
  \frac1{[j - \sigma [j] ]} \;\;\;.
\eeqn
  Putting eqs.~(\ref{eq:cont1g})  and (\ref{eq:cont2g}) together,
  we find
\beqn
\label{eq:***}
 &~&  \left( \begin{array}{cccccc}
 m_{1}, & \cdots, & m_{k}, & 0,     & \cdots, & 0 \\
 1,     & \cdots  & k    , & k+1, & \cdots, & n+1
 \end{array} \right)_{n+1}    \;\;\;  \nonumber \\
 &=&   -  \sum_{\ell}^{k}  \sum_{\sigma \in {\cal S}_{n} }
 \frac1{[1-\sigma(1)]^{m_{1}+1}} \cdots
 \frac1{[\ell-\sigma(\ell)]^{m_{\ell}}} \;
 \frac{  \left \{ 1-  \left( \frac{[\ell - \sigma(\ell)]}{[\ell- (n+1)]}
 \right)^{m_{\ell}}
 \right \}  } { [\sigma(\ell)- n] [\ell -n] }  \nonumber \\
 &\times&  \frac1{[ (\ell +1) -\sigma(\ell +1)]^{m_{\ell +1}+1}} \cdots
 \frac1{[k- \sigma(k)]^{m_{k}+1}}
 \prod_{j ( \neq 1, 2, \cdots k)}^{n-1}
 \frac1{[j - \sigma(j)]} \;\;\;.
\eeqn
  Factorizing the expression inside the bracket, we have
\beqn
\label{eq:recursion2}
 && \left( \begin{array}{cccccc}
 m_{1}, & \cdots, & m_{k}, & 0,     & \cdots, & 0 \\
 1,     & \cdots  & k    , & k+1    & \cdots, & n+1
 \end{array} \right)_{n+1}    \;\;\;  \nonumber \\
 &&\qquad =\sum_{j=1}^{k}\sum_{l=1}^{m_j}
 \left( \begin{array}{cccccccc}
 m_{1}, & \cdots, & m_{j}-l, & \cdots, & m_{k}, & 0,     & \cdots, & 0 \\
 1,     & \cdots  & j,       & \cdots, & k    , & k+1    & \cdots, & n
 \end{array} \right)_{n}
 \frac{1}{[j-n]^{1+l} }
 \quad . \nonumber \\
\eeqn
Then from the assumption eq.~(\ref{eq:relation3}) at $n$,
 eq.~(\ref{eq:relation3})
 in which $n$ is replaced by $n+1$ is also satisfied.
On the other hand for $n=3$  eq.~(\ref{eq:relation3}) is clearly satisfied, so
 we have proven eq.~(\ref{eq:relation3}) and eq.~(\ref{eq:relation4}) .


\end{document}